\documentclass[12pt]{article}
\usepackage{times}   
\usepackage[singlelinecheck=false]{caption} 
\usepackage{graphicx}

\usepackage{xfrac}
\usepackage{relsize}
\usepackage{amsmath}    
\usepackage{cite}   
\usepackage{sectsty}
\subsubsectionfont{\large}
\begin{document}

%
%
%
\bigskip \textbf{ \LARGE \\ Experimental proposal for the Dynamical Spacetime approach to wavefunction collapse
\\
\\
}

%
%
%

\noindent \textit{Garrelt Quandt-Wiese}
\footnote{
\scriptsize My official last name is Wiese. For non-official concerns, my wife and I use our common family name: Quandt-Wiese.}

\noindent \textit{\small{Schlesierstr.\,16, 64297 Darmstadt, Germany}}\\ 
\noindent  {\it {\small garrelt@quandt-wiese.de}}\\
\noindent  {\it ~ {\small http://www.quandt-wiese.de}}

%
%
%
\bigskip
\begin{quote}
{\small An experiment for checking the Dynamical Spacetime approach to wavefunction collapse is proposed. The Dynamical Spacetime approach predicts deviations from Born's rule, when a solid evolves into a three-state superposition, and when the displacement between the superposed states is, at the reduction point in time, significantly larger than the spatial variation of the solids nuclei, being typically on the order of a tenth of an \r{A}ngstr\"om. The solid is brought into the three-state superposition by splitting a photon into three beams and by detecting it in each beam by avalanche photodiodes, which displace the solid at different distances with the help of a piezoactuator. The challenge of the experiment is the precise prediction of the setup's reduction point in time to ensure a sufficient separation between the states at this point in time. This is addressed by avoiding interactions of the setup with the environment during superposition, and by a precise calculation of the setup's reduction point in time with the help of a formulary for the Di\'{o}si-Penrose criterion for solids in quantum superpositions. Since the measurement of reduction probabilities is not disturbed by state decoherence, the experiment can be performed at room temperature. The quantitative analysis demonstrates that the predicted increase of the reduction probability of one state by a factor of 1.5 with respect to Born's rule can be measured by a few hundred statistically significant measurements.}
\end{quote}

\bigskip
%
%
%
\section{Introduction}                 
%
%
\label{sec:1}
All models for wavefunction collapse have so far been unable to be checked by experiments. The gravity-based approaches of Di\'{o}si and Penrose \cite{Dio-1,Pen-1} or dynamical reduction models \cite{GRW_Ue-2} require the measurement of the lifetimes of quantum superpositions, which e.g. can be carried out by measuring the vanishing of quantum interference between superposed states, such as in the famous mirror-experiment of Marshall \cite{GExp-8}. However, such procedures are always disturbed by the unavoidable decoherence between the superposed states due to environmental interaction, whose suppression by e.g. very low temperatures or an ultra-high vacuum is difficult to arrange. 

\bigskip
\noindent
A basically new perspective for checking a collapse model is provided by the recently published \textit{Dynamical Spacetime approach} to wavefunction collapse \cite{NS,P2}, which predicts, beside lifetimes of superpositions, new effects in the form of deviations from Born's rule for special regimes. The Dynamical Spacetime approach is as the approaches of Di\'{o}si and Penrose a gravity-based approach, which enhances semiclassical gravity by postulating that the spacetime region on which quantum fields exist and on which the wavefunction's evolution can be regarded is bounded towards the future by a spacelike hypersurface, which is dynamically expanding towards the future. Collapse is displayed in the way that the wavefunction's evolution becomes unstable at certain critical expansions of spacetime, at which it reconfigures via a self-reinforcing mechanism quasi-abruptly to a new evolution resembling a classical trajectory. In EPR experiments, this quasi-abrupt reconfiguration of the wavefunction's evolution can concern far-separated regions. The second important feature of the Dynamical Spacetime approach, which provides the perspective for an experimental check, is its capability to forecast reduction probabilities on the basis of a physical argument. This explains why all experiments performed so far behave in accordance with Born's rule, and predict deviations from it, when solids evolve into three-state superpositions. The Dynamical Spacetime approach is the first collapse model open to deviations from Born's rule, because it does not fear the consequences possibly resulting from superluminal signalling. The explanation of the "spooky action at a distance" by the quasi-abrupt reconfigurations of the wavefunction's evolution does not come in conflict with the principles of relativity \cite{NS,P2}. 

\bigskip
\noindent
Since a measurement of reduction probabilities for proving deviations from Born's rule is not disturbed by state decoherence, one does not have to arrange special experimental conditions, such as low temperatures or an ultra-high vacuum, for supressing it to a minimum. This provides a realistic perspective for checking the Dynamical Spacetime approach; the experiments can even be performed at room temperature!

\bigskip
\noindent
The challenge for checking the Dynamical Spacetime approach follows from the fact that the deviations from Born's rule for solids in three-state superpositions occur only when the displacements between the states are at the reduction point in time, significantly larger than the spatial variation of the solid's nuclei. To ensure this condition, the reduction point time of the setup has to be forecasted as precisely as possible. This is effected with the help of a formulary for the Di\'{o}si-Penrose criterion for solids in quantum superpositions, which is developed in a separate publication \cite{Solid}, but whose study is not essential here. Furthermore, interactions of the setup with the environment during superposition have to be supressed to a minimum, since such interactions influence the reduction point in time. Therefore, the setup during superposition is not in contact with the measuring devices. The experiment's result is determined a sufficient time after reduction, by e.g. connecting a voltmeter with the setup.

\bigskip
\noindent
The remainder of paper is structured as follows. In Sections \ref{sec:2} and \ref{sec:3}, the Dynamical Spacetime approach and the underlying mathematical model are discussed. In Sections \ref{sec:4} and \ref{sec:5}, the setup is introduced, and the experiment's feasibility is demonstrated by a detailed quantitative analysis. In Section \ref{sec:6}, an outlook on pursuing experiments is given.

\newpage
%
%
%
\section{Dynamical Spacetime approach}                 
%
%
\label{sec:2}
In this and the following section, we give a brief overview of the Dynamical Spacetime approach to wavefunction collapse and its mathematical model. For starting with the Dynamical Spacetime approach, the overview article \cite{NS} is recommended. The mathematical model is derived in \cite{P2}.

\bigskip   
\bigskip
\bigskip
\noindent
\textbf{1. Physical approach} \\  
\noindent
The Dynamical Spacetime approach is based on two assumptions: semiclassical gravity and the so-called \textit{Dynamical Spacetime postulate}. 

\bigskip
\noindent
\textit{ \textbf{Semiclassical gravity:}} In semiclassical gravity, the gravitational field is not quantised and spacetime geometry is treated classically \cite{NS-3,NS-4}\footnote{   
The question of whether the gravitational field must be quantised is still the subject of scientific debate \cite{Pen-4,NS-6}, and an issue that has not yet been determined by experiments \cite{NS-15}.
}. As a consequence, superposed states must share the same classical spacetime geometry, even if they prefer (according to general relativity) differently curved spacetimes, which is the case when their mass distributions are different. This provokes a competition between the states for the curvature of spacetime, which is the driver of collapse in the Dynamical Spacetime approach. However, semiclassical gravity alone cannot explain collapse, which is known from studies of the Schr\"odinger-Newton equation to display semiclassical gravity in the Newtonian limit \cite{NS-12,NS-2}. 
    
\bigskip
\noindent
\textit{\textbf{Dynamical Spacetime postulate:}} The Dynamical Spacetime approach postulates that the spacetime region on which quantum fields exist and on which the wavefunction's evolution can be regarded is bounded towards the future by a spacelike hypersurface, the so-called \textit{spacetime border} {\large$\bar{\sigma}$}, which is dynamically propagating towards the future over the so-called \textit{dynamical parameter} {\large$\bar{\tau}$}. The dynamical parameter itself is not an observable quantity (beable), and can be chosen to be dimensionless. This postulate enables a fundamentally new behaviour in the way that the wavefunction's evolution on spacetime can retroactively change to a new evolution, when spacetime expands over {\large$\bar{\tau}$}. This is possible, since the wavefunction's evolution is not governed by the unitary evolution only, but must in addition satisfy a boundary condition on the spacetime border.

\bigskip   
\bigskip
\bigskip
\noindent
\textbf{2. Collapse mechanism} \\  
\noindent
The most important result of the Dynamical Spacetime approach is that it leads to a physical mechanism for collapse. Collapse is displayed in the way that the wavefunction's evolution becomes unstable at certain critical positions of the spacetime border; the so-called \textit{reduction positions} {\large$\bar{\sigma}(\bar{\tau}_{_{C}})$}. At these positions, the wavefunction's evolution reconfigures via a self-reinforcing mechanism quasi-abruptly to a new evolution, which then resembles a classical trajectory. Thereby, spacetime geometry changes in favour of the winning state, which causes the path of the other (competing) state to vanish by simple destructive interference. The lifetimes of superpositions following from this mechanism are identical to those resulting from the gravity-based approaches of Di\'{o}si and Penrose \cite{Dio-1,Pen-1}. 

\bigskip
\noindent
The abrupt reconfigurations of the wavefunction's evolution, which can cover far-distant spacetime regions and which always incorporate the complete history of the superposition (which begins when a wavepacket in configuration space has split), can explain the quantum correlations in EPR experiments, even of such with free choice of measurements. Here Bob can instantaneously influence, by the orientation of his polarisation filter, the polarisation of the photon on Alice's side via the abrupt reconfigurations of the wavefunction's evolution. Although this is a faster-than-light mechanism for the explanation of the "spooky action at a distance", it does not provoke a conflict with relativity, since causality does not evolve along free selectable Lorentz frames in spacetime in the Dynamical Spacetime approach, but is parametrised by the expansion of spacetime (i.e. by the dynamical parameter {\large$\bar{\tau}$}) and evolves quasi-orthogonal to it.

\bigskip   
\bigskip
\bigskip
\noindent
\textbf{3. Reduction probabilities} \\  
\noindent
The second important result of the Dynamical Spacetime approach is its capability to forecast reduction probabilities on basis of a physical argument. The probabilities with which the wavefunction's evolution reconfigures at the reduction positions to one state of the superposition depend on how frequently the intensities of the states fluctuate for decay. This can be expressed in terms of so-called \textit{decay-trigger rates} of the states, which depend on the energy increases the states suffer due to the sharing of spacetime geometry in semiclassical gravity. In the Newtonian limit, these energy increases follow from the fact that the states do not reside in their own gravitational potential (resulting from their mass distributions) but in the mean gravitational potential of the superposed states. For two-state superpositions, this leads to energy increases of the states that are proportional to the intensity of the respective competing state to which the state will decay, and therefore to reduction probabilities being proportional to the state's intensity, i.e. to Born's rule. 
This derivation of Born's rule for two-state superpositions can be adapted to all experiments conducted so far with the help of a property that these experiments have in common: they lead to never more than two different mass distributions at one location. These mass distributions refer e.g. to the cases that a particle "is", or "is not", detected at the location.

\begin{figure}[t]
\centering
\includegraphics[width=11.5cm]{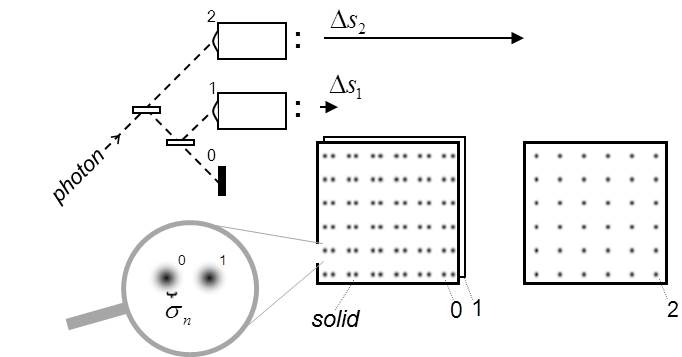}\vspace{0cm}
\caption{\small
Experiment to transfer a solid into a three-state superposition. For photon detection, Detectors 1 and 2 displace the solid by $\Delta s_{1}$, respectively $\Delta s_{2}$.
}
\label{fig1}
\end{figure}

\bigskip   
\bigskip
\bigskip
\noindent
\textbf{4. Deviations from Born's rule for solids in three-state superpositions} \\  
\noindent
The Dynamical Spacetime approach predicts deviations from Born's rule when solids evolve into three-state superpositions. Such a superposition can be created by splitting a photon into three beams and measuring it by detectors, which displace a solid at different distances for photon detection, as shown in Figure \ref{fig1}. When the displacement $\Delta s_{1}$ between the solid in State 0 and State 1 is, at the reduction position, much larger than the spatial variation of its nuclei $\sigma_{n}$, so that the mass distributions of States 0 and 1 are disjoint (cf. Figure \ref{fig1}), and when the displacement between States 2 and 0 is much larger than that between States 1 and 0 ($\Delta s_{2}$$>>$$\Delta s_{1}$), the reduction probability of State 2 increases with respect to Born's rule. Disjoint mass distributions are the criterion that the decay-trigger rates of States 1 and 2 are decorrelated. Then, they both trigger a reconfiguration of the superposition in favour of State 2, which increases the reduction probability of this state with respect to Born's rule.

%

\newpage
%
%
%
\section{Mathematical model}                 
%
%
\label{sec:3}
In this section, we recapitulate the mathematical model of the Dynamical Spacetime approach, which is derived in \cite{P2}. We focus on what is needed for the discussion of the later experiments and limit ourselves to the Newtonian limit. In Section \ref{sec:3.1}, we discuss the basic concepts of the Dynamical Spacetime approach, the so-called \textit{classical scenarios} and \textit{competition actions}, with which we formulate the mathematical model in Section \ref{sec:3.2}. In Section \ref{sec:3.3}, we present the most important formulae of the formulary for the Di\'{o}si-Penrose criterion for solids in quantum superpositions developed in \cite{Solid}, which is needed for the quantitative analysis of the experiments. In Section \ref{sec:3.4}, we apply the model to the experiment in Figure \ref{fig1}, and derive first results for the later discussion.

\bigskip
\bigskip
%
%
%
\subsection{Classical scenarios and competition actions}                 
%
%
\label{sec:3.1}
In this section, we show how the wavefunction's evolution can be decomposed into so-called \textit{classical scenarios}, with which we can conveniently describe the abrupt reconfigurations of the wavefunction's evolution at the critical positions of spacetime border. Furthermore, we introduce the so-called \textit{competition actions}, with which we can measure how much the preferred spacetime geometries of the classical scenarios differ, and how strongly they compete for spacetime geometry.

\bigskip   
\bigskip
\bigskip
\noindent
\textbf{Aligning spacetime border's propagation with the experiment's rest frame} \\  
\noindent
Most predictions of the Dynamical Spacetime approach are fortunately not sensitive to the concrete propagation of the spacetime border $\bar{\sigma}(\bar{\tau})$. The discussion of the experiments in this paper can be simplified by assuming that the spacetime border propagates in coincidence with the experiment's rest frame. The spacetime border is then given by a plane hypersurface, which is specified by a point in time $\bar{t}$ in this rest frame; and the dynamical parameter $\bar{\tau}$ can be expressed by this point in time $\bar{t}$ (i.e. $\bar{\tau}$$\rightarrow$$\bar{t}$). This is very convenient for analyses, since spacetime then simply ends at $\bar{t}$. Accordingly, the reduction positions of spacetime border $\bar{\tau}_{_{C}}$ can be expressed by the corresponding \textit{reduction points in time} $\bar{t}_{_{C}}$.

\bigskip   
\bigskip
\bigskip
\noindent
\textbf{Classical scenarios} \\  
\noindent
Using the convention of classical scenarios, the state vector's evolution $|\psi (t)$$>$ is decomposed into evolutions {\small$|\tilde{\psi}_{_{i}}(t)$$>$} resembling approximately classical trajectories of the system, the \textit{classical scenarios}, as:

\begin{equation}
\label{eq:1}
|\psi(t)>=\sum_{i} c_{_{i}} |\tilde{\psi}_{_{i}} (t)>
\textrm{\textsf{~~,~~~~~~~~~~~}}
\end{equation}

\begin{figure}[t]
\centering
\includegraphics[width=16cm]{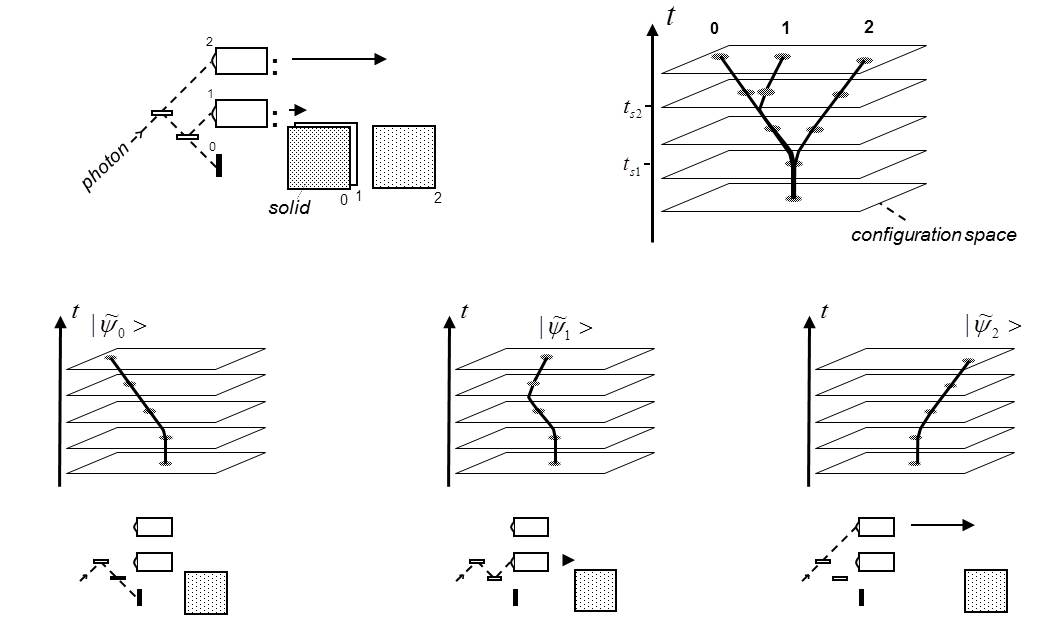}\vspace{0cm}
\caption{\small
Three classical scenarios $|\tilde{\psi}_{_{0}}(t)$$>$, $|\tilde{\psi}_{_{1}}(t)$$>$ and $|\tilde{\psi}_{_{2}}(t)$$>$ (middle part) of the experiment in Figure \ref{fig1} (upper left), which are defined by following up state vector's evolution $|\psi (t)$$>$ on classical paths in configuration space (upper right), and for which the system evolves on classical trajectories in spacetime, as illustrated at the bottom.
}
\label{fig2}
\end{figure}

~
\newline
\noindent with {\small$<$$\tilde{\psi}_{_{i}}(t)|\tilde{\psi}_{_{i}}(t)$$>$$=$$1$} and $\sum_{\mathsmaller{i}}|c_{_{i}}|^{2}$$=$$1$. In our discussion, the state vector $|\psi$$>$ will always describe the complete system, consisting of the experiment in Figure \ref{fig1} of the photon, the beam splitters, the detectors and the solid. The upper right part of Figure \ref{fig2} illustrates state vector's evolution $|\psi (t)$$>$ in configuration space of this experiment, which consists at the beginning of one wavepacket splitting into separate ones when the photon enters a beam splitter. The middle part in Figure \ref{fig2} shows the three classical scenarios {\small$|\tilde{\psi}_{_{0}}(t)$$>$}, {\small$|\tilde{\psi}_{_{1}}(t)$$>$} and {\small$|\tilde{\psi}_{_{2}}(t)$$>$} of the experiment, which are defined by following up the state vector's evolution on classical paths in configuration space, and for which the system evolves on classical trajectories in spacetime. For e.g. Classical Scenario 2 {\small$|\tilde{\psi}_{_{2}}(t)$$>$}, the photon is completely transmitted at the first beam splitter, and is only detected by Detector 2 displacing the solid by $\Delta s_{2}$, as illustrated in the lower right of Figure \ref{fig2}. The classical scenarios are not solutions of Schr\"odinger's equation at the points in time $t$, at  which the state vector $|\psi (t)$$>$ splits into two wavepackets in configuration space, when the photon enters a beam splitter. To fulfil the decomposition of the state vector's evolution according to Equation (\ref{eq:1}) in the regions where several classical scenarios refer to the same root wavepacket, their phases must be chosen suitably. For e.g. the common root wavepacket of all classical scenarios, their phases must satisfy $|\sum_{i}|c_{_{i}}|e^{i\varphi_{i}}|$$=$$1$.

%

\bigskip
\noindent
The concept of classical scenarios is very important for the Dynamical Spacetime approach, since the abrupt reconfigurations of the wavefunction's evolution at collapse can simply be described by intensity shifts between the scenarios. In the Dynamical Spacetime approach, the intensity of a path in configuration space, such as e.g. the path of State 2 between $t_{_{s1}}$ and $\bar{t}$ in Figure \ref{fig2}, drops when the spacetime geometry changes to the disadvantage of this path. This enforces, due to the norm conservation of unitary evolution, that at the wavepacket's splitting point at $t_{_{s1}}$ more intensity is rerouted to the other path belonging to States 0 and 1. This rerouting of intensity at $t_{_{s1}}$ can be expressed by shifting intensity from Classical Scenario 2 to the Scenarios 0 and 1. Thus, the reconfigurations of the wavefunction's evolution can be described by intensity shifts between the classical scenarios\footnote{   
The intensity shifts between classical scenarios must be accompanied by readjustments of their phases $\varphi_{_{i}}$ in the regions where they refer to common root wavepackets to satisfy Equation (\ref{eq:1}) after reconfiguration.
}. 

\bigskip
\noindent
In the following discussion, we abbreviate the intensities $|c_{_{i}}|^{2}$ of the classical scenarios by

\begin{equation}
\label{eq:2}
I_{_{i}}\equiv |c_{_{i}}|^{2}
\textrm{\textsf{~~.~~~~~~~~~~~~~~\footnotesize \textit{intensities of classical scenarios}}}
\end{equation}

\bigskip   
\bigskip
\bigskip
\noindent
\textbf{Di\'{o}si-Penrose energies and competition actions} \\  
\noindent
In this section, we present the so-called \textit{Di\'{o}si-Penrose energies}, which define a measure of how much the preferred spacetime geometries of two states differ from each other. They coincide with the characteristic gravitational energy resulting from the gravity-based collapse models of Di\'{o}si and Penrose \cite{P1}.

\bigskip
\noindent
How much the preferred spacetime geometries of two states differ on an area $A$, which we call the \textit{bundle area}, can be measured with the help of the so-called \textit{local Di\'{o}si-Penrose energies}, which depend on how much the mass distributions of the states differ from each other on $A$. For the discussion of the experiment in Figure \ref{fig1}, we regard for the local Di\'{o}si-Penrose energies the area of the solid ($A$$=$$S$) and the areas of the two detectors ($A$$=$$D1,D2$). For the discussion of the concrete experiments in Section \ref{sec:5}, we take into account that not only the solid on $A$$=$$S$ has different mass distributions in its states, but that also the two detectors have slightly different mass distributions in their detection and no-detection states. On the bundle area of Detector 2 ($A$$=$$D2$), States 0 and 1, both referring to the no-detection case of this detector, have identical mass distributions, which we combine to a so-called \textit{local bundle} on this area. Accordingly, States 0 and 2 are a local bundle on the area of Detector 1 ($A$$=$$D1$). The local Di\'{o}si-Penrose energies $E^{^{A}}_{G\kappa \nu}$ are defined between two of such local bundles ($\kappa$ and $\nu$) on $A$, where a local bundle can also be a single state. The Di\'{o}si-Penrose energy is given by the integral over the difference of the bundles' mass distributions $\rho_{_{\kappa}}(\mathbf{x})$$-$$\rho_{_{\nu}}(\mathbf{x})$ multiplied by the difference of their gravitational potentials $\Phi_{_{\kappa}}(\mathbf{x})$$-$$\Phi_{_{\nu}}(\mathbf{x})$ resulting from the mass distributions\footnote{\small   
I.e.  $\Phi_{_{\kappa}}(\mathbf{x})=-G\int d^{3}\mathbf{y}\rho_{_{\kappa}}(\mathbf{y}) / |\mathbf{x}-\mathbf{y}|$.
}. On the bundle areas of Detector 1 and 2, we can define respectively one local Di\'{o}si-Penrose energy $E^{^{D1}}_{G}$ and $E^{^{D2}}_{G}$ between the two bundles referring to the detection and no-detection cases of the detector as follows \cite{P1}:

\begin{equation}
\label{eq:3}
\begin{split}
E^{^{Di}}_{G}=\frac{1}{2} \int_{\mathbf{x} \in Di}  d^{3}\mathbf{x}(\rho_{_{det}}(\mathbf{x})-\rho_{_{no-det}}(\mathbf{x}))(\Phi_{_{no-det}}(\mathbf{x})-\Phi_{_{det}}(\mathbf{x}))
\textrm{\textsf{~~.~~~~~~~~}}   \\
\textrm{\textsf{~~~~~~~~~~~~~~~~~~~~~~~~~~~~~~~~~~~~~~~~~~~~~~~\footnotesize \textit{Di\'{o}si-Penrose energies of the detectors}}}
\end{split}
\end{equation}

~
\newline
\noindent Accordingly, we obtain three local Di\'{o}si-Penrose energies between States 0, 1 and 2 on the bundle area of the solid ($E^{^{S}}_{G01}$, $E^{^{S}}_{G02}$, $E^{^{S}}_{G12}$), which are given by

\begin{equation}
\label{eq:4}
\begin{split}
E^{^{S}}_{Gij}=\frac{1}{2} \int_{\mathbf{x} \in S}  d^{3}\mathbf{x}(\rho_{_{i}}(\mathbf{x})-\rho_{_{j}}(\mathbf{x}))(\Phi_{_{j}}(\mathbf{x})-\Phi_{_{i}}(\mathbf{x}))
\textrm{\textsf{~~.~~~~~~~~~~~~~~~~}}   \\
\textrm{\textsf{~~~~~~~~~~~~~~~~~~~~~~~~~~~~~~~~~~~~~~~~~~~~~~~\footnotesize \textit{Di\'{o}si-Penrose energies of the solid}}}
\end{split}
\end{equation}

~
\newline
\noindent How much the preferred spacetime geometries of two local bundles of classical scenarios $\kappa$ and $\nu$ differ on the spacetime region, which is given by the area $A$, and limited towards the future by the spacetime border at $\bar{t}$, is measured by the so-called \textit{local competition actions} $S^{^{A}}_{G\kappa \nu}(\bar{t})$. They are defined by integrating the local Di\'{o}si-Penrose energies between the corresponding bundles of states over time until the spacetime border at $\bar{t}$ ($S^{^{A}}_{G\kappa \nu}(\bar{t})$$=$$\int^{\bar{t}}_{..}$$dt E^{^{A}}_{G\kappa \nu}(t)$). The local competition actions between the detection and no-detection cases of the detectors ($S^{^{D1}}_{G}(\bar{t})$, $S^{^{D2}}_{G}(\bar{t})$) are given by

\begin{equation}
\label{eq:5}
S^{^{Di}}_{G}(\bar{t})= \int^{\bar{t}}_{..} dt E^{^{Di}}_{G}(t)
\textrm{\textsf{~~,~~~~~~\footnotesize \textit{competition actions of the detectors}}}
\end{equation}

~
\newline
\noindent and the local competition actions between the three states on the bundle area of the solid ($S^{^{S}}_{G01}(\bar{t})$, $S^{^{S}}_{G02}(\bar{t})$, $S^{^{S}}_{G12}(\bar{t})$) by

\begin{equation}
\label{eq:6}
S^{^{S}}_{Gij}(\bar{t})= \int^{\bar{t}}_{..} dt E^{^{S}}_{Gij}(t)
\textrm{\textsf{~~.~~~~~~\footnotesize \textit{competition actions of the solid}}}
\end{equation}

\bigskip
\bigskip
%
%
%
\subsection{Reconfiguration equation and rule}                 
%
%
\label{sec:3.2}
In this section, we present the mathematical model of the Dynamical Spacetime approach, which describes when, how and with which probabilities the wavefunction's evolution reconfigures.

\bigskip   
\bigskip
\bigskip
\noindent
\textbf{Reconfiguration equation} \\  
\noindent
The so-called \textit{reconfiguration equation} is a conditional equation that depends on the intensities of the classical scenarios $I_{_{i}}$ and the local competition actions $S^{^{A}}_{G\kappa \nu}(\bar{t})$ between them, and whose solutions determine whether intensity changes of the classical scenarios $dI_{_{i}}$ are possible. It is given by the following set of equations (one equation for every state $i$) \cite{P2}:

\begin{equation}
\label{eq:7}
dI_{_{i}}=\sum_{A;\, \kappa \mathsmaller{\supseteq} i}\frac{dI_{_{i}}}{dI_{_{\kappa}}}\sum_{\nu \neq \kappa}\frac{S^{^{A}}_{G\kappa\nu}(\bar{t})}{\hbar}(I_{_{\nu}}dI_{_{\kappa}}-I_{_{\kappa}}dI_{_{\nu}})
\textrm{\textsf{~~.~~~~~~~\footnotesize \textit{reconfiguration equation}}}
\end{equation}

~
\newline
\noindent Here the outer sum runs over all bundle areas $A$, on which local competition actions $S^{^{A}}_{G\kappa \nu}(\bar{t})$ can be defined. The condition $\kappa$$\mathsmaller{\supseteq}$$i$ selects for an area $A$ the local bundle $\kappa$, which contains the regarded state $i$. The inner sum runs over all other local bundles $\nu$ on $A$ competing with $\kappa$ for spacetime geometry. $I_{_{\kappa}}$ and $I_{_{\nu}}$ are the intensities and $dI_{_{\kappa}}$, $dI_{_{\nu}}$ the intensity changes of the local bundles $\kappa$ and $\nu$, where the intensity of a local bundle is given by the sum of the intensities of its states ({\small$I_{_{\kappa}}$$=$$\sum_{\mathsmaller{i\in \kappa}}I_{_{i}}$}).

\bigskip
\noindent
The reduction point in time $\bar{t}_{_{C}}$ of a superposition of classical scenarios is given by the lowest value of $\bar{t}$ for which the reconfiguration equation has non-vanishing solutions for the intensity changes $dI_{_{i}}$.

\bigskip
\noindent
For our experiment in Figure \ref{fig1}, the reconfiguration equation is given by\footnote{\small  
The term $dI_{_{1}}/dI_{_{det}}S^{^{D1}}_{G}(\bar{t})(I_{_{no-det}}dI_{_{det}}$$-$$I_{_{det}}dI_{_{no-det}})$ occurring for the area of Detector 1 ($A$$=$$D1$) is transformed with $I_{_{det}}$$+$$I_{_{no-det}}$$=$$1$ and $dI_{_{det}}$$+$$dI_{_{no-det}}$$=$$0$ to $dI_{_{1}}S^{^{D1}}_{G}(\bar{t})$, and the terms $dI_{_{i}}/dI_{_{no-det}}S^{^{D1}}_{G}(\bar{t})(I_{_{det}}dI_{_{no-det}}$$-$$I_{_{no-det}}dI_{_{det}})$ with $i$$=$$0,2$ to $dI_{_{i}}S^{^{D1}}_{G}(\bar{t})$. The terms corresponding to Detector 2 are transformed accordingly.
}

\begin{equation}
\label{eq:8}
\left(
\begin{matrix}
\mathsmaller{dI_{_{0}}} \\
\mathsmaller{dI_{_{1}}} \\
\mathsmaller{dI_{_{2}}} 
\end{matrix} 
\right)
=
\mathsmaller{\frac{1}{\hbar}}
\left[ 
\left(
\begin{matrix}
\mathsmaller{S^{S}_{G01}I_{_{1}}+ S^{S}_{G02}I_{_{2}}} & \mathsmaller{-S^{S}_{G01}I_{_{0}}} & \mathsmaller{-S^{S}_{G02}I_{_{0}}}  \\
\mathsmaller{-S^{S}_{G01}I_{_{1}}}  & \mathsmaller{S^{S}_{G01}I_{_{0}}+ S^{S}_{G12}I_{_{2}}}  & \mathsmaller{-S^{S}_{G12}I_{_{1}}}   \\
\mathsmaller{-S^{S}_{G02}I_{_{2}}}  & \mathsmaller{-S^{S}_{G12}I_{_{2}}}  & \mathsmaller{S^{S}_{G02}I_{_{0}}+ S^{S}_{G12}I_{_{1}}}   
\end{matrix} 
\right)
\mathsmaller{+}
(\mathsmaller{\mathsmaller{S^{^{D1}}_{G}}}\mathsmaller{+}\mathsmaller{\mathsmaller{S^{^{D2}}_{G}}})
\left(
\begin{matrix}
\mathsmaller{1} & \mathsmaller{0} & \mathsmaller{0}  \\
\mathsmaller{0} & \mathsmaller{1} & \mathsmaller{0}  \\
\mathsmaller{0} & \mathsmaller{0} & \mathsmaller{1}  
\end{matrix} 
\right)
\right]   
\left(
\begin{matrix}
\mathsmaller{dI_{_{0}}} \\
\mathsmaller{dI_{_{1}}} \\
\mathsmaller{dI_{_{2}}} 
\end{matrix} 
\right)
\textrm{\textsf{.}}
\end{equation}

~
\newline
\noindent The reduction point in time $\bar{t}_{_{C}}$ follows by determining the point in time $\bar{t}$ for which the largest eigenvalue $e_{_{max}}$ of the matrix inside the square bracket reaches Planck's quantum of action, i.e. $e_{_{max}}(\bar{t}_{_{C}})$$=$$\hbar$. The corresponding eigenvector {\small$d\vec{I}$$\equiv$$(dI_{_{0}},dI_{_{1}},dI_{_{2}})^{T}$} is called the \textit{reconfiguration solution} $d\vec{I}_{_{C}}$. The intensity vectors of the classical scenarios  {\small$\vec{I}$$\equiv$$(I_{_{0}},I_{_{1}},I_{_{2}})^{T}$} after reconfiguration of the wavefunction's evolution at collapse are either {\small$\vec{I}'_{_{+}}$$=$$\vec{I}$$+$$\hat{\alpha}_{_{+}} d\vec{I}_{_{C}}$} or {\small$\vec{I}'_{_{-}}$$=$$\vec{I}$$-$$\hat{\alpha}_{_{-}} d\vec{I}_{_{C}}$}, where $\hat{\alpha}_{+}$ and $\hat{\alpha}_{-}$ are the largest numbers for which {\small$\vec{I}$$+$$\hat{\alpha}_{_{+}} d\vec{I}_{_{C}}$} respectively {\small$\vec{I}$$-$$\hat{\alpha}_{_{-}} d\vec{I}_{_{C}}$} have no negative components \cite{P2}. In favour of which final state ($\vec{I}'_{_{+}}$ or $\vec{I}'_{_{-}}$) the wavefunction's evolution reconfigures at $\bar{t}_{_{C}}$ depends on smallest intensity fluctuations, which can be described with the so-called \textit{decay-trigger rates} of the local bundles.

\bigskip   
\bigskip
\bigskip
\noindent
\textbf{Decay-trigger rates} \\  
\noindent
In semiclassical gravity, the energy of a local bundle on a bundle area $A$ increases due to the sharing of spacetime geometry with the other competing bundles $\nu$ on $A$ \cite{P1}. These energy increases divided by Planck's constant $\hbar$ determine the decay-trigger rates $dp_{_{\kappa \downarrow}}/d\bar{t}$ of the local bundles, which describe the probability $dp_{_{\kappa \downarrow}}$ for an intensity fluctuation for decay (i.e. the probability for a decay-trigger) during spacetime border moving by $d\bar{t}$ \cite{P2}. The energy increase $E_{_{G\kappa}}$ of a local bundle $\kappa$ depends on the intensities $I_{_{\nu}}$ of the competing bundles $\nu$ on $A$ multiplied by the Di\'{o}si-Penrose energies $E^{^{A}}_{G\kappa \nu}$ between them, i.e. $E_{_{G\kappa}}$$=$$\sum_{\mathsmaller{\nu \neq \kappa}}$$I_{_{\nu}}E^{^{A}}_{G\kappa \nu}$ \cite{P1}. The decay-trigger rates of the three local bundles on the area of the solid, which are identical to States 0, 1 and 2, are given by

\begin{equation}
\label{eq:9}
\begin{split}
\frac{dp_{_{0\downarrow}}}{d\bar{t}}=\frac{1}{\hbar}(I_{_{1}}E^{^{S}}_{G01}+ I_{_{2}}E^{^{S}}_{G02}) \\
\frac{dp_{_{1\downarrow}}}{d\bar{t}}=\frac{1}{\hbar}(I_{_{0}}E^{^{S}}_{G01}+ I_{_{2}}E^{^{S}}_{G12}) \\
\frac{dp_{_{2\downarrow}}}{d\bar{t}}=\frac{1}{\hbar}(I_{_{0}}E^{^{S}}_{G02}+ I_{_{1}}E^{^{S}}_{G12})
\end{split}
\textrm{\textsf{~~~~.~~~~~~~~~\footnotesize \textit{decay-trigger rates of states}}}
\end{equation}

~
\newline
\noindent Since the two detectors in the experiment in Figure \ref{fig1} will be designed in such a way that their Di\'{o}si-Penrose energies are much smaller than those of the solid ($E^{^{Di}}_{G}$$<<$$E^{^{S}}_{Gij}$; see Section \ref{sec:5.3}), we do not present here the decay-trigger rates of the detectors $dp^{Di}_{det \downarrow}/d\bar{t}$ and $dp^{Di}_{no-det \downarrow}/d\bar{t}$ referring to the detection and no-detection bundles on $A$$=$$D1,D2$. Their impact on the final reduction probabilities can be easily determined with the results derived in \cite{P2}.

\bigskip   
\bigskip
\bigskip
\noindent
\textbf{Reconfiguration rule} \\  
\noindent
In favour of which final state, {\small$\vec{I}'_{_{+}}$$=$$\vec{I}$$+$$\hat{\alpha}_{_{+}} d\vec{I}_{_{C}}$} or {\small$\vec{I}'_{_{-}}$$=$$\vec{I}$$-$$\hat{\alpha}_{_{-}} d\vec{I}_{_{C}}$}, the decay-trigger rates of States 0, 1 and 2 trigger the reconfiguration process depend on the projection of these states on the reconfiguration solution $d\vec{I}_{_{C}}$, which is given by $dI_{_{Ci}}$ \cite{P2}. For $dI_{_{Ci}}$$<$$0$, a decay-trigger of state $i$ triggers a reconfiguration to {\small$\vec{I}'_{_{+}}$$=$$\vec{I}$$+$$\hat{\alpha}_{_{+}} d\vec{I}_{_{C}}$}, and for $dI_{_{Ci}}$$>$$0$ to {\small$\vec{I}'_{_{-}}$$=$$\vec{I}$$-$$\hat{\alpha}_{_{-}} d\vec{I}_{_{C}}$}. This can be summarised by the following so-called \textit{reconfiguration rule}\footnote{\small   
$\Theta_{_{0}}(x)$$=$$0$ for $x$$\leq$$0$, $\Theta_{_{0}}(x)$$=$$1$ for $x$$>$$0$.
} \cite{P2}: 

\begin{equation}
\label{eq:10}
\vec{I}\rightarrow \left\{
\begin{matrix}
\vec{I}'_{_{+}}$$=$$\vec{I}$$+$$\hat{\alpha}_{_{+}} d\vec{I}_{_{C}} 
\textrm{\textsf{~~\footnotesize \textit{with}~~}} 
p_{_{+}}\propto \mathlarger{\sum_{i}}\Theta_{_{0}}(-dI_{_{Ci}}) \mathlarger{\frac{dp_{_{i\downarrow}}}{d\bar{t}}} \\
~ \\
\vec{I}'_{_{-}}$$=$$\vec{I}$$-$$\hat{\alpha}_{_{-}} d\vec{I}_{_{C}}
\textrm{\textsf{~~\footnotesize \textit{with}~~~}}
p_{_{-}}\propto \mathlarger{\sum_{i}}\Theta_{_{0}}( dI_{_{Ci}}) \mathlarger{\frac{dp_{_{i\downarrow}}}{d\bar{t}}}
\end{matrix}
\right.
\textrm{\textsf{~~,~~~~~~~\footnotesize \textit{reconfiguration rule}}}
\end{equation}

~
\newline
\noindent which describes the possible reconfigurations of the intensity vector $\vec{I}$ and the relative probabilities $p_{_{+}}$ and $p_{_{-}}$ of these reconfigurations. The absolute reconfiguration probabilities follow by normalisation, i.e. by $p_{_{+}}$$+$$p_{_{-}}$$=$$1$.

\bigskip   
\bigskip
\bigskip
\noindent
\textbf{Decorrelation criterion} \\  
\noindent
When the displacement between two states becomes very small, such as e.g. between States 0 and 1 in Figure \ref{fig1} for $\Delta s_{1}$$\rightarrow$$0$, their decay-trigger rates are correlated, and cannot be counted twice in the reconfiguration rule. The decay-trigger rates of two states are decorrelated when their mass distributions are disjoint \cite{P2}. For States 0 and 1 in Figure \ref{fig1}, this is given when the displacement between the states $\Delta s_{1}$ is at least six times larger than the spatial variation $\sigma_{n}$ of the solid's nuclei\footnote{\small   
The mass distributions of the solid's nuclei can be described by Gaussian distributions, i.e. by $\rho(\mathbf{x})$$\propto$$exp(-\mathbf{x}^{2}/(2\sigma^{2}_{n}))$ \cite{Solid}.
} (cf. inset in Figure \ref{fig1}). The spatial variation $\sigma_{n}$ of a solid's nuclei is typically on the order of a tenth of an \r{A}ngstr\"om \cite{Solid}. This leads to the following so-called \textit{decorrelation criterion}:

\begin{equation}
\label{eq:11}
\Delta s_{1}(\bar{t}_{_{C}})>6\sigma_{_{n}}
\textrm{\textsf{~~,~~~~~~~~~~~~~~\footnotesize \textit{decorrelation criterion}}}
\end{equation}

~
\newline
\noindent which has to be fulfilled at the reduction point in time $\bar{t}_{_{C}}$ \cite{P2}.

\bigskip
\bigskip
%
%
%
\subsection{Di\'{o}si-Penrose energies of superposed solids}                 
%
%
\label{sec:3.3}
In this section, we present the most important formulae of the formulary for the Di\'{o}si-Penrose criterion for solids in quantum superpositions developed in \cite{Solid}, which we need for the quantitative analysis in Section \ref{sec:5}. 

\bigskip
\noindent
The solid that will evolve into a three-state superposition with the setup proposed in Section \ref{sec:4} is a capacitor with a piezo as dielectric, which we call the \textit{piezo capacitor}. The capacitor's plates are pressed apart from each other by the converse piezo electric effect when it is charged. In \cite{Solid}, we developed the basic formulae, with which we could calculate the Di\'{o}si-Penrose energy of such a piezo capacitor in a quantum superposition. The contribution of the capacitor's plates is calculated with a formula developed for a \textit{displaced plate}, which is displaced by $\Delta s$ vertically to its surface in one state of the superposition.  The contribution of the piezo is calculated with a formula developed for an \textit{extended plate}, whose thickness $d$ is changed by $\Delta d$ in one state of the superposition. 

\bigskip
\noindent
For displacements much larger than the mean lattice constant of the solid $\bar{g}$ ($\Delta s$$>>$$\bar{g}$), which is typically in the order of two \r{A}ngstr\"oms ($\bar{g}$$\approx$$2${\footnotesize \AA}), the solid can be approximated by a continuum \cite{Solid}. The Di\'{o}si-Penrose energy resulting from this approximation is called the \textit{long-distance contribution} to the Di\'{o}si-Penrose energy. For smaller displacements, the microscopic mass distribution of the solid's nuclei must be taken into account, which leads to a further contribution: the so-called \textit{short-distance contribution} to the Di\'{o}si-Penrose energy. The Di\'{o}si-Penrose energies of the displaced and extended plate are given by

\begin{equation}
\label{eq:12}
E^{^{S}}_{G}(\Delta s)=
2\pi \alpha_{_{geo}}GV\rho^{2}\Delta s^{2} + E^{^{Ss}}_{G}(\Delta s)
\textrm{\textsf{~~,~~~~~~~~~~~~~~~~~}}
\end{equation}

~
\newline
\noindent where the first term describes the long-distance and $E^{^{Ss}}_{G}(\Delta s)$ the short distance contribution to the Di\'{o}si-Penrose energy. In this result, $G$ is the gravitational constant, $\rho$ the mass density of the solid, $V$ the volume of the plate and $\alpha_{_{geo}}$ the so-called \textit{geometric factor}, which is $1$ for the displaced and $\frac{1}{3}$ for the extended plate. For the extended plate, $\Delta s$ describes the displacement of the plate's surface, i.e. $\Delta s$$=$$\Delta d/2$. The term $E^{^{Ss}}_{G}(\Delta s)$ for the short-distance contribution is given by 

\begin{equation}
\label{eq:13}
E^{^{Ss}}_{G}(\Delta s)
=
\overline{T}^{S}_{G}V\cdot
\left\{   
\begin{matrix}
\mathlarger{\frac{\alpha_{_{geo}}}{12}(\frac{\Delta s}{\sigma_{n}})^{2}} ~~~~~ \Delta s<<\sigma_{n} \\
~ \\
\mathlarger{F_{_{geo}}(\frac{\Delta s}{\sigma_{n}})} ~~~~~~~ \Delta s>4\sigma_{n}
\end{matrix}
\right.  ~~~~,~~~~~~~~~~~~~~~
\end{equation}

~
\newline
\noindent where $\overline{T}^{S}_{G}$ is the so-called \textit{characteristic Di\'{o}si-Penrose energy density of a solid} \footnote{\small   
The characteristic Di\'{o}si-Penrose energy density of a solid is given by 
\newline
\newline
\hspace*{10mm} $\overline{T}^{S}_{G}/\hbar$$=$$\mathlarger{\frac{G\hat{q}\rho^{2}\bar{g}^{3}}{\sqrt{\pi}\sigma_{n}}}$~~ ,
\newline
\newline
where $\hat{q}$ is the so-called \textit{quadratic mass factor}, which is one for solids consisting of only one chemical element \cite{Solid}. 
}, $\sigma_{n}$ the spatial variation of the solid's nuclei and $F_{_{geo}}(x)$ the so-called \textit{geometric function} \footnote{\small   
The geometric functions of the displaced and extended plate are given by \cite{Solid} 
\newline
\newline
\hspace*{10mm} $F_{_{d-pl}}(x)=1-\frac{\sqrt{\pi}}{x}$ ,
\newline
\hspace*{10mm} $F_{_{e-pl}}(x)=1-\frac{2+\sqrt{\pi}/2-\sqrt{\pi}ln(4)}{x} - \mathsmaller{\sqrt{\pi}\frac{ln(x)}{x}}$~.
\newline
}, which converges for $x$$>>$$1$ to one. The characteristic Di\'{o}si-Penrose energy density of a solid divided by Planck's constant $\overline{T}^{S}_{G}/\hbar$ ranges from $4MHz/cm^{3}$ for aluminium, to over $42MHz/cm^{3}$ for iron, and up to $730MHz/cm^{3}$ for iridium \cite{Solid}. The spatial variation $\sigma_{n}$ of a solid's nuclei at room temperature is typically on the order of one-tenth of an \r{A}ngstr\"om ($\sigma_{n}$$\approx$\,$0.1${\footnotesize \AA}\footnote{\small   
At room temperature, the spatial variation of the nuclei is mainly determined by the excited acoustical phonons, and can be estimated with the solid's Debye temperature $\Theta_{_{D}}$ by \cite{Solid}
\newline
\newline
\hspace*{10mm} $\sigma_{_{n}}$$=$$\mathlarger{\sqrt{\frac{3T}{k_{_{B}}\overline{m}}}\frac{\hbar}{\Theta_{_{D}}}}$~~ ,
\newline
\newline
where $k_{_{B}}$ is Boltzmann's constant,  $T$ the temperature and $\overline{m}$ the mean mass of the solid's nuclei.  
}). For displacements much larger than the nuclei's spatial variation ($\Delta s$$>>$$\sigma_{n}$), the short-distance contribution to the Di\'{o}si-Penrose energy converges to a constant value, which is given by $\overline{T}^{S}_{G}V$. This contribution can be neglected in Equation (\ref{eq:12}) for displacements much larger than the solid's mean lattice constant ($\Delta s$$>>$$\bar{g}$).

\bigskip
\bigskip
%
%
%
\subsection{Application}                 
%
%
\label{sec:3.4}
In this section, we apply the mathematical model of the Dynamical Spacetime approach to the experiment in Figure \ref{fig1}. We derive results that we can use for the quantitative discussion of the real experiments in Section \ref{sec:5}.

\bigskip   
\bigskip
\bigskip
\noindent
\textbf{Impact of the detectors} \\  
\noindent
The impact of the two detectors in Figure \ref{fig1} on the reduction point in time $\bar{t}_{_{C}}$ can be discussed with Equation (\ref{eq:8}). When $e^{S}_{max}$ is the largest eigenvalue of the left matrix in Equation (\ref{eq:8}) corresponding to the solid, $\bar{t}_{_{C}}$ is given by

\begin{equation}
\label{eq:14}
e^{S}_{max}(\bar{t}_{_{C}}) +
S^{^{D1}}_{G}(\bar{t}_{_{C}}) +
S^{^{D2}}_{G}(\bar{t}_{_{C}}) = \hbar
\textrm{\textsf{~~.~~~~~~~~~~~~~~\footnotesize \textit{reduction condition}}}
\end{equation}

~
\newline
\noindent This result shows that the competition actions of the detectors $S^{^{D1}}_{G}$ and $S^{^{D2}}_{G}$ shorten the reduction point in time $\bar{t}_{_{C}}$. Too great competition actions of the detectors can therefore prevent the displacement between States 0 and 1 at the reduction point not being sufficiently large to satisfy the decorrelation criterion {\small$\Delta s_{1}(\bar{t}_{_{C}} )$$>$$6 \sigma_{n}$}. In Section \ref{sec:5.3}, we show that the detectors can be designed in such a way that their Di\'{o}si-Penrose energies are much smaller than those of the solid ($E^{^{Di}}_{G}$$<<$$E^{^{S}}_{Gij}$), which allows us to neglect their impact on the reduction point in time. In the following discussion, the competition actions of the detectors are therefore no longer considered.

\bigskip   
\bigskip
\bigskip
\noindent
\textbf{Procedure for calculating $\bar{t}_{_{C}}$ and $p_{_{2}}$} \\  
\noindent
In this section, we describe the general procedure for calculating the reduction point in time $\bar{t}_{_{C}}$  and the reduction probability of State 2 $p_{_{2}}$ of the experiment in Figure \ref{fig1}. This procedure is used for the exact numerical calculations of the experimental proposal in Section \ref{sec:5}. 
When we know the time profiles of the displacements between the states in Figure \ref{fig1} $\Delta s_{i}(t)$, we can determine the Di\'{o}si-Penrose energies of the solid $E^{^{S}}_{Gij}(t)$ with Equation (\ref{eq:12}), and the corresponding competition actions $S^{^{S}}_{Gij}(\bar{t})$ with Equation (\ref{eq:6}).  With the competition actions $S^{^{S}}_{Gij}(\bar{t})$ and the intensity vector $\vec{I}$, we can then calculate the largest eigenvalue $e^{S}_{max}$ of the left matrix in Equation (\ref{eq:8}) as a function of $\bar{t}$, and obtain the reduction point in time $\bar{t}_{_{C}}$ by 

\begin{equation}
\label{eq:15}
e^{S}_{max}(\bar{t}_{_{C}}) = \hbar
\textrm{\textsf{~~.~~~~~~~~~~~~~~\footnotesize \textit{reduction condition}}}
\end{equation}

~
\newline
\noindent With the reconfiguration solution $d\vec{I}_{_{C}}$, i.e. the eigenvector corresponding to $e^{S}_{max}$ at $\bar{t}_{_{C}}$, and the decay-trigger rates $dp_{_{i}}/d\bar{t}$ of the states at the reduction point in time, which can be calculated with the corresponding Di\'{o}si-Penrose energies $E^{^{S}}_{Gij}(\bar{t}_{_{C}})$ (Equation \ref{eq:9}), we obtain with the reconfiguration rule (Equation \ref{eq:10}) the final states {\small$\vec{I}'_{_{+}}$$=$$\vec{I}$$+$$\hat{\alpha}_{_{+}} d\vec{I}_{_{C}}$}, {\small$\vec{I}'_{_{-}}$$=$$\vec{I}$$-$$\hat{\alpha}_{_{-}} d\vec{I}_{_{C}}$} and their reduction probabilities $p_{_{+}}$, $p_{_{-}}$.  In these final states, the experiment can still be in a two-state superposition. Such a two-state superposition will reduce at a later point in time, where the reduction probabilities follow Born's rule. The overall reduction probability for a reduction to State 2 $p_{_{2}}$ is therefore given by

\begin{equation}
\label{eq:16}
p_{_{2}} = p_{_{+}} I'_{_{+2}} + p_{_{-}} I'_{_{-2}}
\textrm{\textsf{~~.~~~~~~~~~~~~~~~~~}}
\end{equation}

\clearpage   
\noindent
\textbf{Case "$\Delta s_{2}$$>>$$\Delta s_{1}$"} \\  
\noindent
In this section, we apply the procedure above to the case that the displacement between States 0 and 2 is much larger than that between States 0 and 1 ($\Delta s_{2}$$>>$$\Delta s_{1}$), as shown in Figure \ref{fig1}. The results of this limiting case can already be used for rough calculations of the real experiments, as will be shown in Section \ref{sec:5}.

\bigskip
\noindent
Since the Di\'{o}si-Penrose energies scale with the square of the displacement between the states (cf. Equation 12), the competition actions, which result from the Di\'{o}si-Penrose energies according to Equation (\ref{eq:6}), can be approximated for $\Delta s_{2}$$>>$$\Delta s_{1}$ by

\begin{equation}
\label{eq:17}
\begin{split}
S^{^{S}}_{G02}(\bar{t}) \approx S^{^{S}}_{G12}(\bar{t})  \\
S^{^{S}}_{G01}(\bar{t}) \approx 0 ~~~~~~~~~
\end{split}
\textrm{\textsf{~~.~~~~~~~~~~~~~~~~~}}
\end{equation}

~
\newline
\noindent With the reduction condition $e^{S}_{max}(\bar{t}_{_{C}})$$=$$\hbar$, follows then

\begin{equation}
\label{eq:18}
S^{^{S}}_{G02} (\bar{t}_{_{C}})=\hbar
 \textrm{\textsf{~~.~~~~~~~~~~~~~~\footnotesize \textit{reduction condition}}}
\end{equation}

~
\newline
\noindent When the displacements (and therefore also the Di\'{o}si-Penrose energies) between the states are constant over time, which means that the competition action   is given by $S^{^{S}}_{G02}(\bar{t})$$=$$E^{^{S}}_{G02}\bar{t}$, this reduction condition leads to $\bar{t}_{_{C}}$$=$$\,\hbar/E^{^{S}}_{G02}$. This reduction point in time $\bar{t}_{_{C}}$ coincides with the lifetimes of superpositions predicted by the gravity-based approaches of Di\'{o}si and Penrose \cite{Dio-3,Pen-1}.

\bigskip
\noindent
The reconfiguration solution $d\vec{I}_{_{C}}$ corresponding to $e^{S}_{max}(\bar{t}_{_{C}})$ is (cf. Equation \ref{eq:8}):

\begin{equation}
\label{eq:19}
d\vec{I}_{_{C}} =
\left(
\begin{matrix}
-I_{_{0}} \\
-I_{_{1}} \\
I_{_{0}} + I_{_{1}}
\end{matrix}
\right)
\textrm{\textsf{~~.~~~~~~~~~~~~~~~~~}}
\end{equation}

~
\newline
\noindent This leads, with the reconfiguration rule (Equation \ref{eq:10}), to the final states:

\begin{equation}
\label{eq:20}
\vec{I}'_{_{+}} = \vec{I}+\hat{\alpha}_{_{+}}d\vec{I}_{_{C}}=
\left(
\begin{matrix}
0 \\
0 \\
1
\end{matrix}    
\right)
= \vec{I}_{_{2}} ~~,
~~~~~~
\vec{I}'_{_{-}} = \vec{I}-\hat{\alpha}_{_{-}}d\vec{I}_{_{C}}=
\mathsmaller{\frac{1}{I_{0}+I_{1}} }
\left(
\begin{matrix}
I_{_{0}} \\
I_{_{1}} \\
0
\end{matrix}    
\right)
= \vec{I}_{_{01}} 
\textrm{\textsf{~~,~~}}
\end{equation}

~
\newline
\noindent where $\vec{I}'_{_{+}}$ corresponds to State 2, which we abbreviated $\vec{I}_{_{2}}$, and $\vec{I}'_{_{-}}$ to a superposition of States 0 and 1, which we abbreviated $\vec{I}_{_{01}}$. 
The decay-trigger rates of States 0 and 1 are both $I_{_{2}}\cdot E^{^{S}}_{G12}/\hbar$, which follows with $E^{^{S}}_{G02}$$\approx$$E^{^{S}}_{G12}$, $E^{^{S}}_{G01}$$\approx$$0$ and Equation (\ref{eq:9}), and trigger according to the reconfiguration rule (Equation \ref{eq:10}) a reconfiguration to $\vec{I}_{_{2}}$($dI_{_{C0}}$$<$$0$, $dI_{_{C1}}$$<$$0$), when the decorrelation criterion {\small$\Delta s_{1}(\bar{t}_{_{C}} )$$>$$6 \sigma_{n}$} is satisfied. The decay-trigger rate of State 2 is $(I_{_{0}}$$+$$ I_{_{1}})\cdot E^{^{S}}_{G12}/\hbar$ and triggers a reconfiguration to $\vec{I}_{_{01}}$ ($dI_{_{C2}}$$>$$0$). This leads to $p_{_{2}}$$\propto$$2I_{_{2}}$, $p_{_{01}}$$\propto$$(I_{_{0}}$$+$$I_{_{1}})$ and the following increased reduction probability of State 2 with respect to Born's rule ($p_{_{2}}$$=$$I_{_{2}}$):

\begin{equation}
\label{eq:21}
p_{_{2}}=\frac{2}{1+I_{_{2}}} I_{_{2}}
\textrm{\textsf{~~~~~~~~\footnotesize \textit{for}~~~~~~}}   
\Delta s_{1}(\bar{t}_{_{C}})>6\sigma_{_{n}}
\textrm{\textsf{~~.~~~~~~~~~~~}}
\end{equation}

~
\newline
\noindent For  $\Delta s_{1}$$=$$0$, when the decay-trigger rates of States 0 and 1 are correlated and account only once in the reconfiguration rule, we obtain $p_{_{2}}$$\propto$$I_{_{2}}$, $p_{_{01}}$$\propto$$(I_{_{0}}$$+$$I_{_{1}})$ and a reduction probability of State 2 in accordance with Born's rule:

\begin{equation}
\label{eq:22}
p_{_{2}} = I_{_{2}}
\textrm{\textsf{~~~~~~~~~~\footnotesize \textit{for}~~~~~~}}   
\Delta s_{1} = 0
\textrm{\textsf{~~.~~~~~~~~~~~}}
\end{equation}

\bigskip   
\bigskip
\bigskip
\noindent
\textbf{Case "$\Delta s_{2}/\Delta s_{1}$$=$$4$"} \\  
\noindent
We now calculate a case that is already fairly close to the real experiments that will be discussed in Section \ref{sec:5}. We regard the case that the displacement between States 0 and 2 is for all times $t$ four times larger than that between States 0 and 1, i.e. $\Delta s_{2}(t)/\Delta s_{1}(t)$$=$$4$, and that the displacements are much larger than the solid's mean lattice constant ($\Delta s_{i}$$>>$$\bar{g}$). The following calculations will show that the results do not differ much from the previous case "$\Delta s_{2}$$>>$$\Delta s_{1}$". This means that Equations (\ref{eq:18}) and (\ref{eq:21}) can be used for rough calculations of $\bar{t}_{_{C}}$ and $p_{_{2}}$ of the real experiments.

\bigskip
\noindent
According to Equations (\ref{eq:12}) and (\ref{eq:13}), the Di\'{o}si-Penrose energies between the three states $E^{^{S}}_{Gij}$ scale with the displacements between them, which are given by $\Delta s_{i}$$-$$\Delta s_{j}$\footnote{\small   
The displacement $\Delta s_{0}$ is defined by $\Delta s_{0}$$\equiv$$0$.
}, like $E^{^{S}}_{Gij}(t)$$\propto$$(\Delta s_{i}$$-$$\Delta s_{j})^{2}$, when the displacements are much larger than the solid's mean lattice constant ($\Delta s_{i}$$>>$$\bar{g}$). When the displacement ratio $\Delta s_{2}/\Delta s_{1}$ is constant over time, the ratios between the Di\'{o}si-Penrose energies and the competition actions are given for all time $t$ by the ratios between the squared displacements $(\Delta s_{i}$$-$$\Delta s_{j})^{2}$ (e.g. $S^{^{S}}_{G12}(\bar{t})/S^{^{S}}_{G02}(\bar{t})$$=$$3^{2}/4^{2}$). This simplifies the calculation of the largest eigenvalue $e^{S}_{max}$ of the left matrix in Equation (\ref{eq:8}) as a function of $\bar{t}$. For an intensity vector of {\small$\vec{I}$$=$$(\frac{1}{2},\frac{1}{4},\frac{1}{4})^{T}$}, the reduction condition $e^{S}_{max}(\bar{t}_{_{C}})$$=$$\hbar$ leads to 

\begin{equation}
\label{eq:23}
S^{S}_{G02} (\bar{t}_{_{C}}) = 1.15 \hbar
 \textrm{\textsf{~~,~~~~~~~~~~~~~~\footnotesize \textit{reduction condition}}}
\end{equation}

~
\newline
\noindent which does not differ much from Equation (\ref{eq:18}) for "$\Delta s_{2}$$>>$$\Delta s_{1}$". The reconfiguration solution $d\vec{I}_{_{C}}$ corresponding to $e^{S}_{max}(\bar{t}_{_{C}})$ is

\begin{equation}
\label{eq:24}
d\vec{I}_{_{C}} =
\left(
\begin{matrix}
-0.626 \\
-0.140 \\
0.767
\end{matrix}
\right)  
\textrm{\textsf{~~.~~~~~~~~~~~~~~~~~}}
\end{equation}

~
\newline
\noindent With the reconfiguration rule (Equation \ref{eq:10}), we obtain for {\small$\Delta s_{1}(\bar{t}_{_{C}} )$$>$$6 \sigma_{n}$} the following final states and corresponding reduction probabilities: 

\begin{equation}
\label{eq:25}
\begin{split}
\vec{I}'_{_{+}}=\vec{I}+\hat{\alpha}_{_{+}} d\vec{I}_{_{C}} = 
\left(
\begin{matrix}
0 \\
0.138 \\
0.862
\end{matrix}
\right)  
\textrm{\textsf{~~\footnotesize \textit{with}~~}} 
p_{_{+}}=0.406 
\\
~ \\
\vec{I}'_{_{-}}=\vec{I}-\hat{\alpha}_{_{-}} d\vec{I}_{_{C}} =
\left(
\begin{matrix}
0.704 \\
0.296 \\
0
\end{matrix}
\right)  
\textrm{\textsf{~~\footnotesize \textit{with}~~~}}
p_{_{-}}=0.594
\end{split}
\textrm{\textsf{~~~~~.~~~~~~~~~~~~~~~~~}}
\end{equation}

\bigskip
\noindent
The overall reduction probability of State 2 follows with Equation (\ref{eq:16}) as $p_{_{2}}$$=$$0.35$, which is by a factor of $1.40$ larger than that expected by Born's rule ($p_{_{2}}$$=$$I_{_{2}}$$=$$\frac{1}{4}$). With Equation (\ref{eq:21}) for "$\Delta s_{2}$$>>$$\Delta s_{1}$", we obtain with {\small$I_{_{2}}=\frac{1}{4}$} an increase of  $p_{_{2}}/I_{_{2}}$$=$$1.6$ with respect to Born's rule, which does not differ much from $p_{_{2}}/I_{_{2}}$$=$$1.40$. The result of our calculation can be summarised by 

\begin{equation}
\label{eq:26}
\begin{matrix}
S^{S}_{G02}(\bar{t}_{_{C}}) = 1.15 \hbar \\
~~~ p_{_{2}} / I_{_{2}} = 1.40
\end{matrix}
\textrm{\textsf{~~~~~\footnotesize \textit{for}~~}} 
\left\{
\begin{matrix}
\Delta s_{2}(t) / \Delta s_{1}(t) = 4  \\
\Delta s_{2}(\bar{t}_{_{C}}) >> \bar{g} ~~~~~~~~ \\
\Delta s_{1}(\bar{t}_{_{C}})>6\sigma_{_{n}}  ~~~~~~ \\
\vec{I} = (\frac{1}{2}, \frac{1}{4}, \frac{1}{4})^{T} ~~~~~~~~
\end{matrix}  
\right.
\textrm{\textsf{~~.~~~~~~~}}
\end{equation}

\newpage
%
%
%
\section{Experiment}                 
%
%
\label{sec:4}
In this section, we present the experiment for checking deviations from Born's rule. In Section \ref{sec:4.1}, we introduce the setup. In Section \ref{sec:4.2}, we show how the interaction of the setup with the environment is minimised during superposition. In Section \ref{sec:4.3}, we present the process of a measurement; and in Section \ref{sec:4.4}, we summarise the technical parameters of the photodiodes that are needed for the quantitative analysis in Section \ref{sec:5}.

%
\begin{figure}[h]
\centering
\includegraphics[width=12cm]{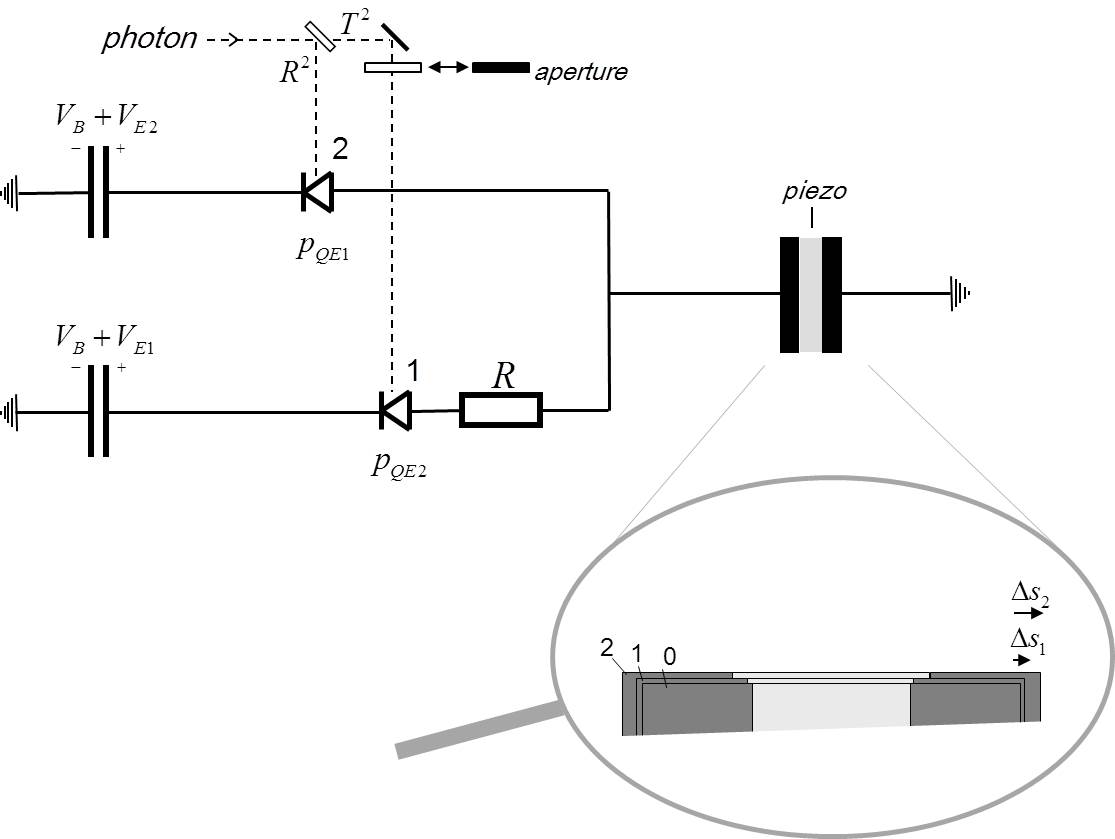}\vspace{0cm}
\caption{\small
Setup to allow the capacitor on the right to evolve into a three-state superposition, as illustrated in the inset.
}
\label{fig3}
\end{figure}


%
%
%
\subsection{Setup}                 
%
%
\label{sec:4.1}
Figure \ref{fig3} shows the setup to allow solid to evolve into a three-state superposition. The solid that will evolve into this superposition is the capacitor at the right with the piezo as dielectric, which we call the \textit{piezo capacitor}. The capacitor's plates are pressed apart from each other by the converse piezoelectric effect when it is charged, i.e. the piezo capacitor is working like a piezoactuator. The piezo capacitor is transferred into a three-state superposition by splitting the photon with the beam splitter into two beams and by measuring it with the two avalanche photodiodes in Figure \ref{fig3}, whose avalanche currents charge the capacitor with different strength, and which let the capacitor's plates be displaced by displacements of $\Delta s_{2}$ in State 2, and $\Delta s_{1}$ in State 1, as shown in the inset. The displacements $\Delta s_{1}$ and $\Delta s_{2}$ can be controlled by the photodiode's excess bias voltages $V_{_{E1}}$ and $V_{_{E2}}$, and the resistor $R$ behind Photodiode 1. The third state, i.e. State 0, in which the capacitor's plates are not displaced at all, occurs due to the finite detection probabilities of the photodiodes (being smaller than one) automatically, which are called the photodiodes' \textit{quantum efficiencies} $p_{_{QE1}}$ and $p_{_{QE2}}$. The intensities of States 0, 1 and 2 of the so-generated three-state superposition depend on the photodiode's quantum efficiencies $p_{_{QE1}}$ and $p_{_{QE2}}$, and the transmission coefficient $T^{2}$ and reflection coefficient $R^{2}$ of the beam splitter ($T^{2}$$+$$R^{2}$$\approx$$1$) as follows:

\begin{equation}
\label{eq:27}
\begin{split}
I_{_{0}} = 1- T^{2}p_{_{QE1}} - R^{2}p_{_{QE2}} \\
I_{_{1}} =  T^{2}p_{_{QE1}} ~~~~~~~~~~~~~~~~~~~~~ \\
I_{_{2}} =  R^{2}p_{_{QE2}}  ~~~~~~~~~~~~~~~~~~~~~
\end{split}
\textrm{\textsf{~~.~~~~~~~~~~~~~~~~~}}
\end{equation}

\bigskip   
\bigskip
\noindent
\textbf{Quantum mechanical origin of quantum efficiencies} \\  
\noindent
We assume that the photodiodes' finite quantum efficiencies ($p_{_{QE1}},$ $p_{_{QE2}}$$<$$1$) have a quantum mechanical origin. This means that the photodiodes evolve after the photon's arrival into a superposition of a detection and no-detection state, reducing to one of these states after some time. If the photodiodes' finite quantum efficiencies would have a classical origin, such as e.g. thermal fluctuations, we could not use them for steering the intensities of the states according to Equation (\ref{eq:27}). In this case, the photodiodes have to be operated with as large as possible quantum efficiencies, and the three-state superposition has to be generated by splitting the photon into three beams, as in Figure \ref{fig1}.

\bigskip
\bigskip
%
%
%
\subsection{Minimising setup's interaction with the environment}                 
%
%
\label{sec:4.2}
To observe deviations from Born's rule, we have to ensure that the displacement between States 0 and 1 is at the reduction point in time sufficiently large to fulfil the decorrelation criterion {\small$\Delta s_{1}(\bar{t}_{_{C}} )$$>$$6 \sigma_{n}$}. Since the reduction point in time cannot be measured so far, we must forecast it as precisely as possible to ensure a sufficient displacement at this point in time. This is effected on the one hand by a precise calculation, taking all components of the setup into account (Section \ref{sec:5}); and on the other hand, by minimising the setup's interaction with the environment during superposition. The second measure ensures that the environment, such as e.g. an observer, does not participate in quantum superposition and influence the reduction point in time. The minimisation of environmental interaction is addressed in our experiment by the following two measures.

\bigskip   
\bigskip
\bigskip
\noindent
\textbf{1. Measurement after reduction instead of reduction by measurement} \\  
\noindent
Since an observer, who tries to measure the result of the experiment directly, participates in quantum superposition and thus influences the reduction point in time, we take the result of our experiment a sufficient period of time after the three-state superposition of the solid has reduced. This is carried out by connecting after this period of time voltmeters to the two capacitors in Figure \ref{fig3}, which are biasing Photodiode 1 and 2, and by checking whether their voltages have dropped due to an avalanche current in the corresponding photodiode. A voltage drop in Capacitor 2 corresponds to a reduction to State 2; a drop in Capacitor 1 to a reduction to State 1; and no drop to a reduction to State 0.

\bigskip   
\bigskip
\bigskip
\noindent
\textbf{2. Plate capacitors as voltage supplies} \\  
\noindent
Since the usual voltage supplies can interact with the environment up to the power plant, our avalanche photodiodes are not biased by such voltage supplies. Instead we use, as shown in Figure \ref{fig3}, simple plate capacitors, which are charged before the measurement by the usual voltage supplies, and which are disconnected from the voltage supplies shortly before the photon's arrival.

\bigskip
\bigskip
%
%
%
\subsection{Measurement}                 
%
%
\label{sec:4.3}
In this section, we present the process of measurement.

\bigskip   
\bigskip
\noindent
\textbf{Gated mode} \\  
\noindent
Avalanche photodiodes can have breakthroughs even in the absence of photons. The probability of such \textit{dark counts} increases with the photodiode's excess bias voltage $V_{_{E}}$, i.e. how much the photodiode is biased above its breakdown voltage $V_{_{B}}$. To avoid dark counts, the photodiodes can be operated in the so-called \textit{gated mode} \cite{SPAD-2}. In the gated mode, the photodiodes are biased at the operating level $V_{_{B}}$$+$$V_{_{E}}$ only for a short period of time around the photon's arrival. Before the photon's arrival, we bias the photodiodes slightly below their breakdown voltages $V_{_{B}}$. After the photon's arrival, we disconnect the photodiodes from their biasing plate capacitors, as will be explained subsequently.

\begin{figure}[t]
\centering
\includegraphics[width=14cm]{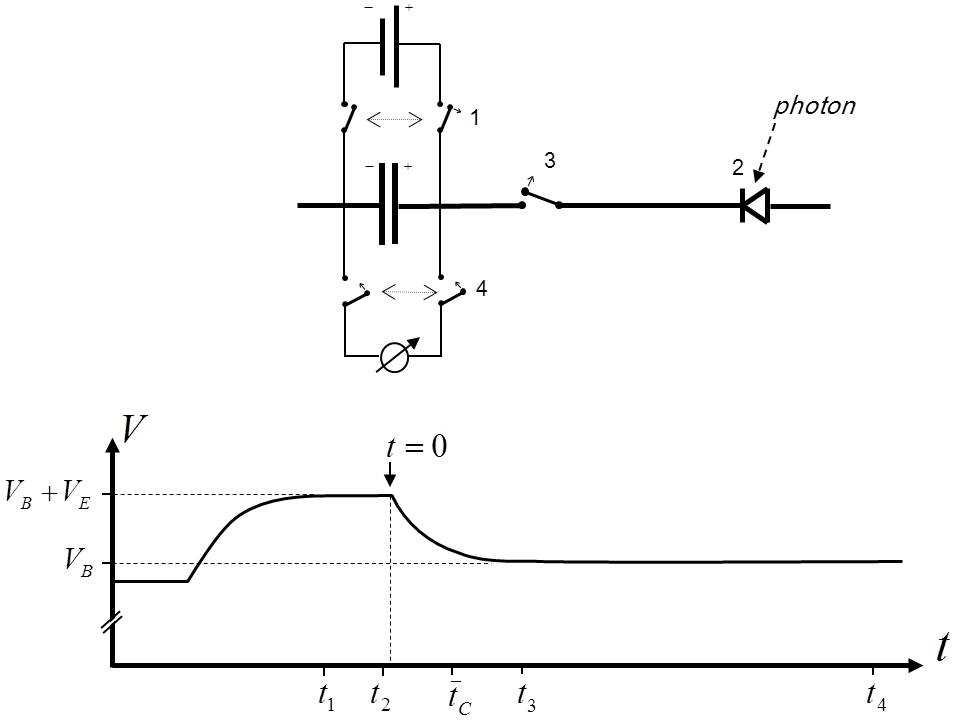}\vspace{0cm}
\caption{\small
\textit{Upper:} \,Setup for charging the plate capacitors and for measuring the voltages at the plate capacitors after the photon's arrival.\\
\textit{Lower:} \,Voltage curve at the photodiode and points in time $t_{_{i}}$ corresponding to steps of the measurement.
}
\label{fig4}
\end{figure}

\bigskip   
\bigskip
\bigskip
\noindent
\textbf{Process} \\  
\noindent
The process of a measurement is illustrated in Figure \ref{fig4}. The upper part of the figure shows the setup for charging the plate capacitors and for measuring the voltages at the plate capacitors after the photon's arrival. The lower part shows the voltage curve at the photodiode during a measurement and points in time $t_{_{i}}$ corresponding to steps of the measurement that will be explained later. At the beginning, the voltage supply is connected to the capacitor by closing the switches at 1, and the voltage at the capacitor is kept slightly below the photodiode's breakdown voltage $V_{_{B}}$. At $t_{_{1}}$, shortly before the photon's arrival, the voltage is increased to $V_{_{B}}$$+$$V_{_{E}}$, and the voltage supply is then disconnected from the capacitor by opening the switches at 1. At $t_{_{2}}$, the photon enters the photodiode, and a short time afterwards the avalanche current starts at $t$$=$$0$. At $t_{_{3}}$, after the avalanche current has stopped, the photodiode is disconnected to suppress dark counts from the plate capacitor by opening the switch at 3. At $t_{_{4}}$, the voltmeter is connected to the capacitor by closing the switches at 4, and the result of the measurement is taken by checking whether the capacitor's voltage has dropped due to an avalanche current. The voltage drop is roughly $\Delta V$$\approx$$V_{_{E}}\cdot C_{_{p}}/(C$$+$$C_{_{p}})$, where $C$ and $C_{_{p}}$ are the capacitances of the plate and the piezo capacitor, respectively.
The four switches for disconnecting the voltage supply and the voltmeter from the setup during superposition in Figure \ref{fig4} can be realised by field effect transistors of enhancement mode. For the execution of the measurement process, it is recommended to use computer-controlled voltage supplies and meters.

\bigskip   
\bigskip
\bigskip
\noindent
\textbf{Reference measurement behaving in accordance with Born's rule} \\  
\noindent
To check the increase of State 2's reduction probability with respect to Born's rule, we make a reference measurement, in which the reduction probability of State 2 behaves in accordance with Born's rule. This is effected by inserting an aperture before Photodiode 1, as shown in Figure \ref{fig3}. The photon is then measured by Photodiode 2 only, and the piezo capacitor evolves into a two-state superposition for which the reduction probability of State 2 behaves in accordance with Born's rule ($p_{_{2}}$$=$$I_{_{2}}$).

%

\bigskip
\bigskip
%
%
%
\subsection{Photodiodes}                 
%
%
\label{sec:4.4}
The proposed experiment can be realised with commercially available so-called \textit{thick silicon SPADs} (single-photon avalanche photodiodes). In this section, we summarise the typical parameters of these photodiodes \cite{SPAD-1} that are used for the quantitative analysis in Section \ref{sec:5}.

\bigskip
\noindent
\textbf{Photon's wavelength:} Thick silicon SPADs are suitable for photon wavelength of $\lambda$$\approx$$800nm$, for which they have quite good quantum efficiencies.

\bigskip
\noindent
\textbf{Breakdown voltage:} The breakdown voltages of thick silicon SPADs are in the range between $250V$-$450V$. In our calculations, we assume $V_{_{B}}$$=$$\,420V$.

\bigskip
\noindent
\textbf{Excess bias voltage:}  Thick silicon SPADs can be operated with excess bias voltages $V_{_{E}}$ in the range of $1V$-$50V$.

\bigskip
\noindent
\textbf{Quantum efficiency:}  The SPADs' quantum efficiencies $p_{_{QE}}$ increase with the applied excess bias voltage $V_{_{E}}$. In our calculations, we assume a quantum efficiency of $p_{_{QE}}$$=$$70\%$ for $V_{_{E}}$$=$$\,20V$, and a quantum efficiency of $p_{_{QE}}$$=$$35\%$ for $V_{_{E}}$$=$$\,10V$.

\bigskip
\noindent
\textbf{Dark count rate:}  The photodiode's dark count rate $f_{_{DC}}$ defining the rate of dark counts in the absence of photons increases also with the excess bias voltages $V_{_{E}}$, and is smaller than $20kHz$ for $V_{_{E}}$$=$$\,20V$.

\bigskip
\noindent
\textbf{Time resolution:}  The time resolution $\Delta t_{_{res}}$ achieved in photon timing decreases over the excess bias voltages, and is typically $170ps$ for $V_{_{E}}$$=$$\,20V$.

\bigskip
\noindent
\textbf{Latching current level:}  The so-called latching current level, below which the avalanche current breaks down, is typically $I_{_{q}}$$\approx$$0.1mA$. 

\bigskip
\noindent
\textbf{Resistance:}  The internal resistance of a thick silicon SPAD is typically lower than $500\Omega$. In our calculations, we assume $R_{_{d}}$$=$$500\Omega$.

\newpage
%
%
%
\section{Quantitative analysis}                 
%
%
\label{sec:5}
In this section, we calculate how the setup's components must be dimensioned and the operating parameters to be chosen to observe deviations from Born's rule. These calculations will demonstrate the feasibility of the experiment. In Section \ref{sec:5.1}, we discuss the setup in Figure \ref{fig3}, in which the three-state superposition is generated with the help of the piezo capacitor. In Section \ref{sec:5.2}, we discuss an alternative, in which a capacitor with movable plates is used for this purpose. In Section \ref{sec:5.3}, we investigate how much the other components of the setup, such as the photodiodes, plate capacitors etc., influence the reduction point in time, and show that their impact is negligible if they are chosen suitably.

\bigskip
\bigskip
%
%
%
\subsection{Setup with piezo capacitor}                 
%
%
\label{sec:5.1}
The quantitative analysis of the setup with the piezo capacitor in Figure \ref{fig3}  is performed in three steps. In Section \ref{sec:5.1.1}, we derive formulae to calculate the setup's reduction point in time. In Section \ref{sec:5.1.2}, we show how the setup must be dimensioned and the operating parameters must be chosen. In Section \ref{sec:5.1.3} we present the results of our calculations, which will demonstrate the feasibility of the proposed experiment.

%
\begin{figure}[h]
\centering
\includegraphics[width=5cm]{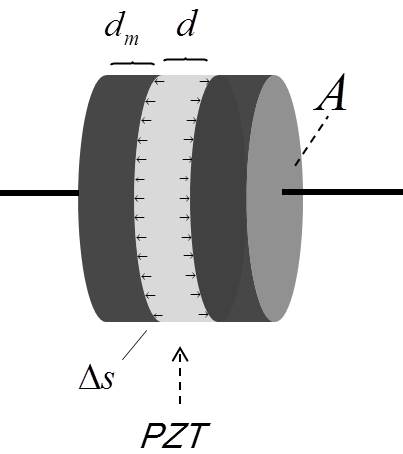}\vspace{0cm}
\caption{\small
Piezo capacitor.
}
\label{fig5}
\end{figure}

\bigskip
%
%
%
\subsubsection{Calculation of reduction point in time}                 
%
%
\label{sec:5.1.1}
Figure \ref{fig5} shows details of the piezo capacitor. It has an area of $A$, its piezo a thickness of $d$, and its plates thicknesses of $d_{_{m}}$.

\bigskip   
\bigskip
\bigskip
\noindent
\textbf{Piezos of PZT} \\  
\noindent
To obtain sufficient displacements of the piezo capacitor's plates for fulfilling the decorrelation criterion {\small$\Delta s_{1}(\bar{t}_{_{C}} )$$>$$6 \sigma_{n}$}, we require piezos with particularly large piezo electric coefficients, given for piezoelectric ceramics of lead zirconium titanate (PZT). The calculations here refer to the product PIC-153 of PI Ceramic GmbH, which is used for piezoactuators \cite{Gen-11b}. For PIC-153, only the $d_{_{33}}$-component of the matrix for the converse piezoelectric effect is relevant\footnote{\small   
The $d_{_{31}}$- and the $d_{_{15}}$-components of PIC-153 are not specified \cite{Gen-11b}.
}, which describes how much the piezo's thickness in the z-direction $d$ changes when an electric field $E$ in the same direction is applied, i.e. $\Delta d/d$$=$$d_{_{33}}E$ \cite{Gen-11b}. PIC-153 has an extremely high $d_{_{33}}$-coefficient (also denoted as the piezoelectric charge coefficient) of $d_{_{33}}$$=$$600\cdot 10^{-10}V^{-1}cm$, accompanied by a very high relative permittivity of $\epsilon_{_{r}}$$=$$4200$ \cite{Gen-11b}\footnote{\small   
An alternative to  PIC-153 is PIC-152 with $d_{_{33}}$$=$$300\cdot 10^{-10}V^{-1}cm$ and $\epsilon_{_{r}}$$=$$1350$, for which also only the $d_{_{33}}$-component is relevant \cite{Gen-11b}.
}. When a voltage of $V_{_{i}}$ is applied to the piezo capacitor, its plates are displaced by

\begin{equation}
\label{eq:28}
\Delta s_{i} = \frac{d_{_{33}}}{2} V_{_{i}}
\textrm{\textsf{~~~,~~~~~~~~~~~~~~~~~}}
\end{equation}

~
\newline
\noindent which follows with $V$$=$$Ed$ and $\Delta s$$=$$\Delta d/2$, where $E$ is the electric field inside the piezo. Equation (\ref{eq:28}) describes the displacement of a piezoactuator with one layer \cite{Gen-11}.

\bigskip   
\bigskip
\bigskip
\noindent
\textbf{Displacement profiles $\Delta s_{i}(t)$} \\  
\noindent
When the capacitances $C$ of the plate capacitors biasing the photodiodes are much larger than that of the piezo capacitor $C_{_{p}}$ ($C$$>>$$C_{_{p}}$), the voltage profiles at the piezo capacitor in States 1 and 2, $V_{_{1}}(t)$ and $V_{_{2}}(t)$ are given by \cite{SPAD-1}: 

\begin{equation}
\label{eq:29}
V_{_{i}}(t) = V_{_{Ei}} (1 - e^{-\mathlarger{\frac{t}{\tau_{_{i}}}}})
\textrm{\textsf{~~~~~\footnotesize \textit{with}~~~~}}
\tau_{_{1}} = (R + R_{_{d}}) C_{_{p}}, ~~~
\tau_{_{2}} = R_{_{d}} C_{_{p}}
\textrm{\textsf{~~,~~~~~~~~~~~~~~~~~}}
\end{equation}

~
\newline
\noindent where $V_{_{E1}}$ and $V_{_{E2}}$ are the excess bias voltages applied to the photodiodes (see Figure \ref{fig3}), $R_{_{d}}$ the internal resistance of the photodiode, and $R$ the resistance of the resistor behind Photodiode 1 in Figure \ref{fig3}. The capacitance of the piezo capacitor is given by $C_{_{p}}$$=$$\epsilon_{_{0}}\epsilon_{_{r}}A/d$. The displacement profiles $\Delta s_{1}(t)$ and $\Delta s_{2}(t)$ of the piezo capacitor's plates in State 1 and 2 follow by inserting Equation (\ref{eq:29}) into Equation (\ref{eq:28}). When the piezo capacitor is completely charged in State 2, its plates are displaced by 

\begin{equation}
\label{eq:30}
\Delta s_{2_{max}} = \frac{d_{_{33}} V_{_{E2}}}{2} 
\textrm{\textsf{~~.~~~~~~~~~~~~~~~~~}}
\end{equation}

\bigskip   
\bigskip
\bigskip
\noindent
\textbf{Di\'{o}si-Penrose energies and competition actions of the piezo capacitor} \\  
\noindent
The Di\'{o}si-Penrose energies of the piezo capacitor between the states $E^{^{S}}_{Gij}$ can be calculated with Equation (\ref{eq:12}), where the contribution of the capacitor's plates is calculated with a geometric factor of $\alpha_{_{geo}}$$=$$1$ (displaced plate), and the contribution of the piezo with $\alpha_{_{geo}}$$=$$\frac{1}{3}$ (extended plate). In \cite{Solid}, it is shown that the Di\'{o}si-Penrose energy of the piezo capacitor is given by the sum of the Di\'{o}si-Penrose energies of the piezo and of the two plates, when the size of the plates is much larger than the thicknesses of the piezo and the plates ($\sqrt{A}$$>>$$d,d_{_{m}}$). For displacements much larger than the mean lattice constant of the solid ($\Delta s_{i}$$>>$$\bar{g}$), for which the short-distance contribution to the Di\'{o}si-Penrose energy $E^{^{Ss}}_{G}(\Delta s)$ in Equation (\ref{eq:12}) can be neglected, the piezo capacitor's total Di\'{o}si-Penrose energies is given by

\begin{equation}
\label{eq:31}
E^{^{S}}_{Gij}(t) = 2 \pi GA (\mathsmaller{\frac{1}{3}} d \rho^{2}_{p} + 2d_{_{m}}\rho^{2}_{m}) \,
(\Delta s_{i}(t)- \Delta s_{j}(t)) ^{2}
\textrm{\textsf{~~,~~~~~~~~~~~~~~~~~}}
\end{equation}

~
\newline
\noindent where $\rho_{_{p}}$ and $\rho_{_{m}}$ are the mass densities of the piezo and the plates. The displacement profile in State 0 is $\Delta s_{0}(t)$$=$$0$. The competition actions of the piezo capacitor $S^{^{S}}_{Gij}(\bar{t})$ (from Equation \ref{eq:6}) are

\begin{equation}
\label{eq:32}
S^{^{S}}_{Gij}(\bar{t})= \int^{\bar{t}}_{0} dt E^{^{S}}_{Gij}(t)
\textrm{\textsf{~~,~~~~~~~~~~~~~~~~~}}
\end{equation}

~
\newline
\noindent where $t$$=$$0$ corresponds to the point in time when the avalanche is triggered (cf. Figure \ref{fig4}).

\bigskip   
\bigskip
\bigskip
\noindent
\textbf{Estimation of $\bar{t}_{_{C}}$} \\  
\noindent
The reduction point in time $\bar{t}_{_{C}}$  of the piezo capacitor in the three-state superposition can be estimated according to the calculations in Section \ref{sec:3.4} for the case "$\Delta s_{2}$$>>$$\Delta s_{1}$" and also the more realistic case "$\Delta s_{2}/\Delta s_{1}$$=$$4$" by $S^{^{S}}_{G02}(\bar{t}_{_{C}})$$\approx$$\hbar$, i.e. when the competition action between Classical Scenarios 0 and 2 approximately reaches Planck's quantum of action. For the calculation of $\bar{t}_{_{C}}$, we have to distinguish the cases where the charging of the piezo capacitor has finished before $\bar{t}_{_{C}}$, or will finish after $\bar{t}_{_{C}}$. When the charging has finished much before $\bar{t}_{_{C}}$, the displacement profile in State 2 can be approximated by $\Delta s_{2}(t)$$\approx$$\Delta s_{2_{max}}$. When it finishes long after $\bar{t}_{_{C}}$, we can assume that $\Delta s_{2}(t)$$\approx$$\Delta s_{2_{max}}\cdot t/\tau_{_{2}}$ (cf. Equations \ref{eq:29} and \ref{eq:30}). The area $A$ of our piezo capacitor, for which the capacitor's charging finishes approximately at the reduction point in time, i.e. $\bar{t}_{_{C}}$$\approx$$\,2\tau_{_{2}}$, is:

\begin{equation}
\label{eq:33}
A_{max} =
\sqrt{\frac{9 \hbar}{\pi G \epsilon_{_{0}}\epsilon_{_{r}}(1+6\frac{d_{m}\rho^{2}_{m}}{d\rho^{2}_{p}}) R_{_{d}} }} \,
\frac{1}{\rho_{_{p}}d_{_{33}}V_{_{E2}}}
\textrm{\textsf{~~,~~~~~~~~~~~~~~~~~}}
\end{equation}

~
\newline
\noindent which follows with the approximation $\Delta s_{2}(t)$$\approx$$\Delta s_{2_{max}}\cdot t/\bar{t}_{_{C}}$ for the displacement profile. For the case $A$$<$$A_{max}/4$ (i.e. when the charging has finished before $\bar{t}_{_{C}}$), $A$$\approx$$A_{max}$ (when the charging finishes at roughly $\bar{t}_{_{C}}$), and $A$$>$$2A_{max}$ (when the charging finishes after $\bar{t}_{_{C}}$), the reduction points in time are given by:

\begin{equation}
\label{eq:34}
\bar{t}_{_{C}} \approx
\left\{
\begin{matrix}
\mathlarger{\frac{6\hbar}
{\pi Gd\rho^{2}_{p}(1+6\frac{d_{m}\rho^{2}_{m}}{d\rho^{2}_{p}})d^{^{2}}_{33}V^{^{2}}_{E2}A}}
~~~ A <  \mathlarger{\frac{A_{max}}{4}}
\\
~ \\
\mathlarger{\sqrt{\frac{9\hbar\epsilon_{_{0}}R_{_{d}}\epsilon_{_{r}}}
{\pi G (1+6\frac{d_{m}\rho^{2}_{m}}{d\rho^{2}_{p}})
}} \, \,
\frac{1}{d\rho_{_{p}}d_{_{33}}V_{_{E2}}}}
~~~ A \approx  A_{max}
\\
~ \\
\mathlarger{\sqrt[3]{ \frac{18\pi \epsilon^{^{2}}_{0}\epsilon^{^{2}}_{r} R^{^{2}}_{d} A}
{\pi G \rho^{^{2}}_{p} (1+6\frac{d_{m}\rho^{2}_{m}}{d\rho^{2}_{p}}) d^{^{2}}_{33} V^{^{2}}_{E2}  }} 
\, \, \frac{1}{d}  }
~~~ A >  2 A_{max}
\end{matrix}  
\right.
\textrm{\textsf{~~.~~~~~}}
\end{equation}

~
\newline
\newline
\noindent The corresponding displacements $\Delta s_{2}$ of the piezo capacitor's plates in State 2 at these points in time are 

\begin{equation}
\label{eq:35}
\Delta s_{2}(\bar{t}_{_{C}}) \approx
\left\{
\begin{matrix}
\mathlarger{\frac{d_{_{33}}V_{_{E2}}}{2}}
~~~~~~~~~~~~~~~~~~~~~~~~~~~~~~~~~~~~~ A \leq  A_{max}
\\
~ \\
\mathlarger{\sqrt[3]{ \frac{9\pi V_{_{E2}} d_{_{33}} }
{4 \pi G \epsilon_{_{0}} \epsilon_{_{r}} \rho^{^{2}}_{p} (1+6\frac{d_{m}\rho^{2}_{m}}{d\rho^{2}_{p}}) R_{_{d}} A^{^{2}}}  } }
~~~ A >  2 A_{max}
\end{matrix}  
\right.
\textrm{\textsf{~~.~~~~~}}
\end{equation}

~
\newline
\noindent The largest possible displacement of the capacitor's plates is $\Delta s_{2_{max}}$$=$$d_{_{33}}V_{_{E2}}/2$ (Equation \ref{eq:30}), which we obtain for $A$$\leq$$A_{max}$. This means that the area $A_{max}$ according to Equation (\ref{eq:33}) is the largest possible area of our piezo capacitor with which we can achieve the maximum displacement $\Delta s_{2_{max}}$. This maximum displacement $\Delta s_{2_{max}}$ and the area $A_{max}$ both depend on parameters of the photodiode, i.e. the excess bias voltage $V_{_{E2}}$ and the photodiode's internal resistance $R_{_{d}}$ (see Equations \ref{eq:30} and \ref{eq:33}). One can excise this dependency by charging the piezo capacitor not with the photodiode's avalanche current, but, as shown in Figure \ref{fig9}, by a separate plate capacitor, which is connected to the piezo capacitor with the help of a field effect transistor whose gate is steered by the photodiode's avalanche current. The excess bias voltage $V_{_{E2}}$ and the internal resistance $R_{_{d}}$ in Equations (\ref{eq:30}) and (\ref{eq:33})-(\ref{eq:35}) then have to be replaced by the voltage $V_{_{2}}$ of the separate plate capacitor and the internal resistance of the field effect transistor connecting the plate capacitor with the piezo capacitor (see Figure \ref{fig9}).

\bigskip
\bigskip
%
%
%
\subsubsection{Choice of parameters}                 
%
%
\label{sec:5.1.2}
In this section, we show how the setup must be dimensioned and the operating parameters must be chosen to observe deviations from Born's rule. For the experimenter, it is important to note that the calculation of the operating parameters and of the dimensioning of the piezo capacitor can be performed with the approximation formulae derived here, and do not require numerical calculations. These are Equations (\ref{eq:21}) and (\ref{eq:27}) for the calculation of the expected increase of State 2's reduction probability with respect to Born's rule $p_{_{2}}/I_{_{2}}$, and Equations (\ref{eq:33})-(\ref{eq:35}) for the calculation of the reduction point in time $\bar{t}_{_{C}}$ and the displacement $\Delta s_{2}(\bar{t}_{_{C}} )$ in State 2 at this point in time.

\bigskip
\noindent
\textbf{Excess bias voltages:} We size our experiment such that we obtain a displacement ratio of about $\Delta s_{2}/\Delta s_{1}$$\approx$$4$ at the reduction point in time, which is (according to our calculations in Section \ref{sec:3.4}) sufficient to obtain a significant increase of State 2's reduction probability with respect to Born's rule. For higher displacement ratios, it is more difficult to satisfy the decorrelation criterion {\small$\Delta s_{1}(\bar{t}_{_{C}} )$$>$$6 \sigma_{n}$}. To obtain in State 2 a larger displacement than in State 1 ($\Delta s_{2}$$>$$\Delta s_{1}$), we choose the excess bias voltage of Photodiode 2 with $V_{_{E2}}$$=$$\,20V$ to be larger than that of Photodiode 1 with $V_{_{E1}}$$=$$\,10V$, which leads, with $\Delta s_{i}$$\approx$$d_{_{33}}V_{_{Ei}}/2$ (cf. Equations \ref{eq:28} and \ref{eq:30}), to a displacement ratio of about $\Delta s_{2}/\Delta s_{1}$$\approx$$2$\footnote{   
Note that a lower excess bias voltage of Photodiode 1 of e.g. only $V_{_{E1}}$$=$$\,5V$ leads to a lower quantum efficiency of this photodiode. 
}. To obtain the displacement ratio of $\Delta s_{2}/\Delta s_{1}$$\approx$$4$, we additionally insert the resistor $R$ behind Photodiode 1 (see Figure \ref{fig3}), whose resistance will be calculated below.

\bigskip
\noindent
\textbf{Beam splitter:} From Equation (\ref{eq:21}), it follows that the absolute increase of State 2's reduction probability with respect to Born's rule $p_{_{2}}$$-$$I_{_{2}}$ is highest for an intensity $I_{_{2}}$ of State 2 on the interval $[0.25,0.6]$. The intensity of State 2 depends on the reflection coefficient $R^{2}$ of the beam splitter and the quantum efficiency $p_{_{QE2}}$ of Photodiode 2 as $I_{_{2}}$$=$$R^{2}p_{_{QE2}}$, where the quantum efficiency $p_{_{QE2}}$ is determined by the photodiode's excess bias voltage $V_{_{E2}}$ (cf. Section \ref{sec:4.4}). The reflection coefficient $R^{2}$ of the beam splitter should therefore be chosen in such a way that $I_{_{2}}$ is on the interval $[0.25,0.6]$. The intensities $I_{_{0}}$ and $I_{_{1}}$ of States 0 and 1 have, according to Equation (\ref{eq:21}), no influence on the increase of State 2's reduction probability and do not require special attention. The reason for this is explained in \cite{P2,NS}.

\bigskip
\noindent
For our calculations, we choose $R^{2}$$=$$30\%$ and $T^{2}$$=$$\,70\%$, which leads with the quantum efficiencies $p_{_{QE2}}$$=$$70\%$ and $p_{_{QE1}}$$=$$35\%$ following from the photodiodes' excess bias voltages of $V_{_{E2}}$$=$$\,20V$ and $V_{_{E1}}$$=$$\,10V$ (cf. Section \ref{sec:4.4}) to an intensity vector of {\small$\vec{I}$$=$$(0.545,0.245,$ $0.21)^{T}$} (cf. Equation \ref{eq:27}), which is close to the intensity vector {\small$\vec{I}$$=$$(\frac{1}{2},\frac{1}{4},\frac{1}{4})^{T}$} that was used for the calculation of the case "$\Delta s_{2}/\Delta s_{1}$$=$$4$" in Section \ref{sec:3.4}. The intensity of State 2 is with $I_{_{2}}$$=$$0.21$ close to the interval $[0.25,0.6]$.

\bigskip
\bigskip
\noindent
\textbf{Plates of the piezo capacitor:} According to Equation (\ref{eq:33}), the area $A_{max}$ to achieve the maximum displacement $\Delta s_{2_{max}}$ decreases with the mass density $\rho_{_{m}}$ of the piezo capacitor's plates. Therefore, it is recommended to use plates with a small mass density. For our calculations, we assume plates of aluminium.

\bigskip
\noindent
\textbf{Dimensions of the piezo capacitor:} The area of our piezo capacitor $A$ is chosen to be close to the area $A_{max}$ for achieving the maximum displacement $\Delta s_{2_{max}}$ according to Equation (\ref{eq:33}). For the calculation of $A_{max}$, we assume that the thickness of the piezo $d$ is twice as large as those of the plates $d_{_{m}}$ ($d$$=$$2d_{_{m}}$). From Equation (\ref{eq:33}) for $A_{max}$, we get with the mass densities of PZT and aluminium of $\rho_{_{PZT}}$$=$$7.6g/cm^{3}$ and $\rho_{_{Al}}$$=$$2.7g/cm^{3}$, the parameters of PIC-153 (cf. Section \ref{sec:5.1.1}), the excess bias voltage of Photodiode 2 of $V_{_{E2}}$$=$$\,20V$, and an internal resistance of the photodiode of $R_{_{d}}$$=$$500\Omega$, an area $A_{max}$, which corresponds to a disc with a diameter of $2.4mm$. For our calculations, we assume a PZT disc with a diameter of $3mm$ and a thickness of $d$$=$$0.2mm$, which is the smallest standard dimension offered by PI Ceramic GmbH \cite{Gen-11b}. The thicknesses of the aluminium plates are $d_{_{m}}$$=$$d/2$$=$$0.1mm$. For these parameters, our piezo capacitor has a capacitance of $C_{_{p}}$$=$$1300pF$.

\bigskip
\noindent
\textbf{Resistor behind Photodiode 1:} The resistance of the resistor $R$ behind Photodiode 1 in Figure \ref{fig3} is chosen such that we obtain the displacement ratio of $\Delta s_{2}/\Delta s_{1}$$=$$4$ at the reduction point in time. This leads to the condition $V_{_{2}}(\bar{t}_{_{C}})/ V_{_{1}}(\bar{t}_{_{C}})$$=$$4$\footnote{\small   
I.e. $V_{_{E2}}(1-e^{-\frac{\bar{t}_{C}}{R_{d}C_{p}}})$$=$$4V_{_{E1}}(1-e^{-\frac{\bar{t}_{C}}{(R_{d}+R)C_{p}}})$.
} (cf. Equations \ref{eq:28} and \ref{eq:29}). The reduction point in time $\bar{t}_{_{C}}$ follows with Equation (\ref{eq:34}) for $A$$\approx$$A_{max}$ to be $\bar{t}_{_{C}}$$\approx$$\,0.86\mu s$. This leads with $V_{_{E2}}$$=$$\,20V$, $V_{_{E1}}$$=$$\,10V$, $R_{_{d}}$$=$$500\Omega$ and $C_{_{p}}$$=$$1300pF$ to a resistance of $R$$=$$940\Omega$.

\bigskip
\bigskip
%
%
%
\subsubsection{Feasibility}                 
\begin{figure}[t]
\centering
\includegraphics[width=13cm]{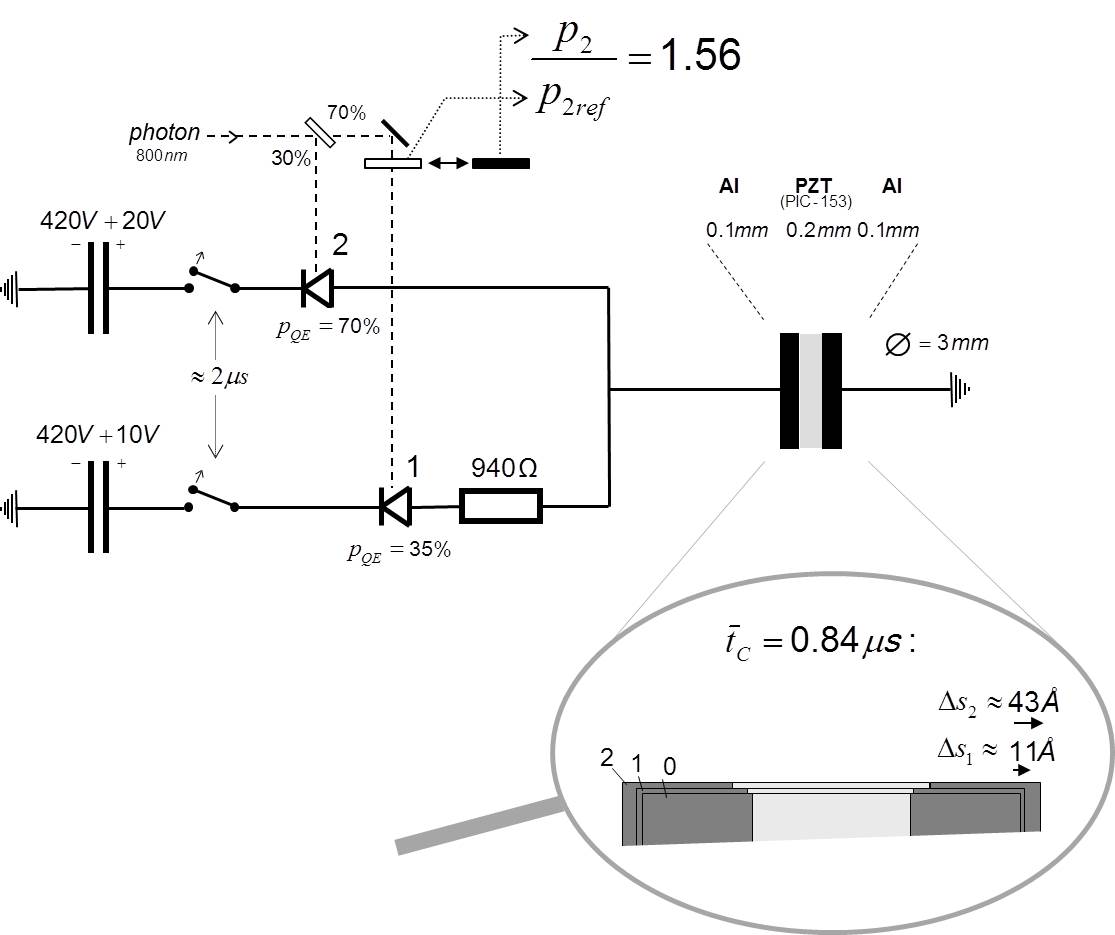}\vspace{0cm}
\caption{\small
Chosen parameters and calculated result of the experiment in Figure \ref{fig3}. The detection probability in Photodiode 2 increases by a factor of $1.56$ when the aperture before Photodiode 1 is removed. The inset shows the displacements of the plates in States 1 and 2 at the reduction point in time $\bar{t}_{_{C}}$.
}
\label{fig6}
\end{figure}

%
%
\label{sec:5.1.3}
In this section, we present the results following from the choice of parameters of Section \ref{sec:5.1.2}, which will demonstrate the feasibility of the experiment. We will calculate the results on the one hand with our approximation formulas and compare them to the exact numerical calculations, which take all discussed details into account. The comparison of the results will show that our approximation formulae are sufficient for the dimensioning of the experiment. Figure \ref{fig6} displays the chosen parameters and the calculated results.

\bigskip
\noindent
\textbf{Rough calculation:} With the approximation formula for $\bar{t}_{_{C}}$ (Equation \ref{eq:34}), we obtain for $A$$\approx$$A_{max}$ a reduction point in time of $\bar{t}_{_{C}}$$=$$\,0.86\mu s$, and with the approximation formula for $p_{_{2}}$ (Equation \ref{eq:21}) with $I_{_{2}}$$=$$0.21$ an increase of State 2's reduction probability with respect to Born's rule of $p_{_{2}}/I_{_{2}}$$=$$1.65$. The displacement in State 2 at the reduction point in time follows with Equation (\ref{eq:35}) for $A$$\leq$$A_{max}$ to be $\Delta s_{2}(\bar{t}_{_{C}} )$$=$$60${\footnotesize \AA}. The displacement in State 1 is, according to our choice of $R$ in Section \ref{sec:5.1.2}, four times smaller; i.e. $\Delta s_{1}(\bar{t}_{_{C}} )$$=$$15${\footnotesize \AA}.

\bigskip
\noindent
\textbf{Exact numerical calculation:} When we calculate $\bar{t}_{_{C}}$ and $p_{_{2}}$ with the procedure described at the beginning of Section \ref{sec:3.4} and calculate the piezo capacitor's Di\'{o}si-Penrose energies with Equation (\ref{eq:31}), in which the short-distance contributions to the Di\'{o}si-Penrose energies are neglected (cf. Section \ref{sec:3.3}), we obtain $\bar{t}_{_{C}}$$=$$\,0.87\mu s$ and $p_{_{2}}/I_{_{2}}$$=$$1.49$. When we take the short-distance contributions to the Di\'{o}si-Penrose energies of the piezo and the plates according to Equation (\ref{eq:13}) additionally into account\footnote{\small   
The short-distance contribution to the Diósi-Penrose energy of the plates is calculated with the geometric function $F_{_{geo}}(x)$ for the displaced plate and $\overline{T}^{S}_{G}/\hbar$$=$$\,4.3MHz/cm^{3}$, $\sigma_{n}$$=$\,$0.1${\footnotesize \AA} for aluminium \cite{Solid}. The short-distance contribution of the piezo is calculated with the geometric function for the extended plate and $\overline{T}^{S}_{G}/\hbar$$=$$\,71.8MHz/cm^{3}$, $\sigma_{n}$$=$\,$0.095${\footnotesize \AA} for PZT \cite{Solid}.
}, we obtain $\bar{t}_{_{C}}$$=$$\,0.84\mu s$, $p_{_{2}}/I_{_{2}}$$=$$1.56$ and displacements of $\Delta s_{2}(\bar{t}_{_{C}} )$$=$$43.3${\footnotesize \AA} and $\Delta s_{1}(\bar{t}_{_{C}} )$$=$$10.7${\footnotesize \AA}, as shown in Figure \ref{fig6}. The short-distance contributions to the Di\'{o}si-Penrose energy do not change the result very much, since the displacements $\Delta s_{2}$ and $\Delta s_{1}$ are significantly larger than the mean lattice constants ($\Delta s_{i}$$>>$$\bar{g}$; cf. Section \ref{sec:3.3}), which are  $\bar{g}$$=$$2.42${\footnotesize \AA} and $\bar{g}$$=$$2.55${\footnotesize \AA} for PZT and aluminium, respectively \cite{Solid}.

%

\bigskip
\noindent
\textbf{Fulfilment of the decorrelation criterion:} The displacement between States 0 and 1 at the reduction point in time is $\Delta s_{1}(\bar{t}_{_{C}} )$$=$$10.7${\footnotesize \AA}; sufficiently large to fulfil the decorrelation criterion {\small$\Delta s_{1}(\bar{t}_{_{C}} )$$>$$6 \sigma_{n}$}. The spatial variation of the nuclei of aluminium and PZT is respectively $\sigma_{n}$$\approx$\,$0.1${\footnotesize \AA} and $\sigma_{n}$$\approx$\,$0.095${\footnotesize \AA} at room temperature \cite{Solid}.

\bigskip
\noindent
\textbf{Breakdown of the avalanche currents:} For the chosen parameters, the three-state superposition of the piezo capacitor reduces before the breakdowns of the avalanche currents in Photodiodes 2 and 1, which happen at $t_{_{q}}$$\approx$$\,4\mu s$ and $t_{_{q}}$$\approx$$\,8\mu s$ when the avalanche currents in Photodiodes 2 and 1 fall below the latching current level of $I_{_{q}}$$\approx$$0.1mA$.

\bigskip
\noindent
\textbf{Settling time of the piezo capacitor:} The settling time for the displacement of the piezo capacitor's plate is smaller than the expected reduction point in time of $\bar{t}_{_{C}}$$\approx$$\,1\mu s$. The typical working frequencies of piezoactuators of PZT reach up to $3MHz$ \cite{Gen-11b}. This means that the time delays occurring between voltage changes at the piezo capacitor and the displacements of its plates can be neglected. The settling time resulting from the finite sound velocities in PZT and in the aluminium plates is $\Delta t$$\approx$$0.05\mu s$\footnote{\small   
This follows with $\Delta t$$\approx$$d/(2v^{^{PZT}}_{||})$$+$$d_{_{m}}/v^{^{Al}}_{||}$ \cite{Solid}  and sound velocities of $v^{^{PZT}}_{||}$$=$$\,2910m/s$ and $v^{^{Al}}_{||}$$=$$\,6420m/s$ for PZT and aluminium, respectively \cite{Solid}.
}. 

\bigskip
\noindent
\textbf{Time resolution of the photodiodes:} The photodiodes' time resolutions are, at typically $170ps$, significantly smaller than the expected reduction point in time, and play no further role in our discussion. 
 
\bigskip
\noindent
\textbf{Dark counts:} From the time constant for charging the piezo capacitor in State 2 of $\tau_{_{2}}$$=$$R_{_{d}}C_{_{p}}$$=$$\,0.65\mu s$ and the reduction point in time of $\bar{t}_{_{C}}$$\approx$$\,1\mu s$, it follows that the photodiodes can be disconnected from their biasing plate capacitors after roughly $2\mu s$, as shown in Figure \ref{fig6}. This leads with $f_{_{DC}}(x)$$<$$20kHz$ for $V_{_{E}}$$=$$\,20V$ (cf. Section \ref{sec:4.4}) to a dark count probability $p_{_{DC}}$ smaller than $0.04$ ($p_{_{DC}}$$=$$f_{_{DC}}\cdot 2\mu s$).
 
\bigskip
\noindent
\textbf{Number of measurements needed:} For the reference measurement at which we insert the aperture before Photodiode 1 in Figure \ref{fig6}, the detection probability in Photodiode 2  behaves in accordance with Born's rule and is $p_{_{2ref}}$$=$$I_{_{2}}$$=$$0.21$. When we remove the aperture, the detection probability in Photodiode 2 increases from $0.21$ to $p_{_{2}}$$=$$I_{_{2}}\cdot 1.56\,$$=$$\,0.33$. From these numbers and a dark count probability $p_{_{DC}}$ smaller than $0.04$, it follows that the predicted increase of Photodiode 2's detection probability of $p_{_{2}}/ p_{_{2 \,ref}}$$=$$1.56$ can be checked by a few hundred statistically significant measurements.

\newpage
%
%
%
\subsection{Setup with a capacitor with movable plates}                 
\begin{figure}[h]
\centering
\includegraphics[width=5cm]{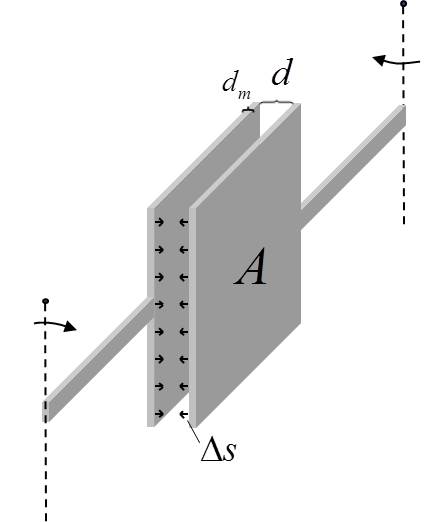}\vspace{0cm}
\caption{\small
Capacitor with movable plates.
}
\label{fig7}
\end{figure}

%
%
\label{sec:5.2}
In this section, we discuss an alternative to the setup in Figure \ref{fig3}, in which we use a capacitor with movable plates (instead of the piezo capacitor) to allow a solid to evolve into a three-state superposition. Such a capacitor is shown in Figure \ref{fig7}, whose plates move towards each other by the electric force between the plates when the capacitor is charged. The capacitor's plates have an area of $A$, a distance of $d$ and thicknesses of $d_{_{m}}$, as shown in Figure \ref{fig7}.

%

\bigskip   
\bigskip
\noindent
\textbf{Reduction point in time} \\  
\noindent
For our choice of parameters, the time constant for charging the capacitor with movable plates will be significantly smaller than the reduction point in time, which allows us to simplify the voltage profiles at the capacitor by $V_{_{i}}(t)$$=$$V_{_{Ei}}$. This leads to the following displacement profiles in States 1 and 2 ($i$$=$$1,2$)\footnote{\small   
This follows with $F$$=$$M\Delta \ddot{s}$, $M$$=$$Ad_{_{m}}\rho_{_{m}}$, $F$$=$$EQ$, $Ed$$=$$V_{_{E}}$, $Q$$=$$V_{_{E}}C$ and $C$$=$$\epsilon_{_{0}}A/d$.
}:

\begin{equation}
\label{eq:36}
\Delta s_{i}(t) =
\frac{\epsilon_{_{0}} V^{^{2}}_{Ei} }
{2 d^{^{2}} d_{_{m}} \rho_{_{m}}  } \,
t^{2}
\textrm{\textsf{~~.~~~~~~~~~~~~~~~~~}}
\end{equation}

~
\newline
\noindent With a calculation similar to that in Section \ref{sec:5.1.1} using Equation (\ref{eq:31}) with $\rho_{_{p}}$$=$$0$ to estimate the Di\'{o}si-Penrose energies $E^{^{S}}_{Gij}$ of the capacitor with movable plates, and with the condition $S^{^{S}}_{G02}(\bar{t}_{_{C}})$$\approx$$\hbar$ to determine the reduction point in time, we obtain the following reduction point in time:

\begin{equation}
\label{eq:37}
\bar{t}_{_{C}} =
\sqrt[5]{
\frac{5\hbar d^{^{4}} d_{_{m}}}
{\pi G \epsilon^{2}_{0} A V^{^{4}}_{E2}  }
}
\textrm{\textsf{~~.~~~~~~~~~~~~~~~~~}}
\end{equation}

~
\newline
\noindent The displacement $\Delta s_{2}$ in State 2 at this point in time is given by

\begin{equation}
\label{eq:38}
\Delta s_{2}(\bar{t}_{_{C}}) =
\sqrt[5]{
\frac{25 \epsilon_{_{0}} \hbar^{2} V^{^{2}}_{E2} }
{32 \pi^{2} G^{^{2}} d^{^{2}} d^{^{3}}_{m} A^{^{2}}}
} \, \,
\frac{1}{\rho_{_{m}}}
\textrm{\textsf{~~~~.~~~~~~~~~~~~~~~~~}}
\end{equation}

%
\begin{figure}[t]
\centering
\includegraphics[width=13cm]{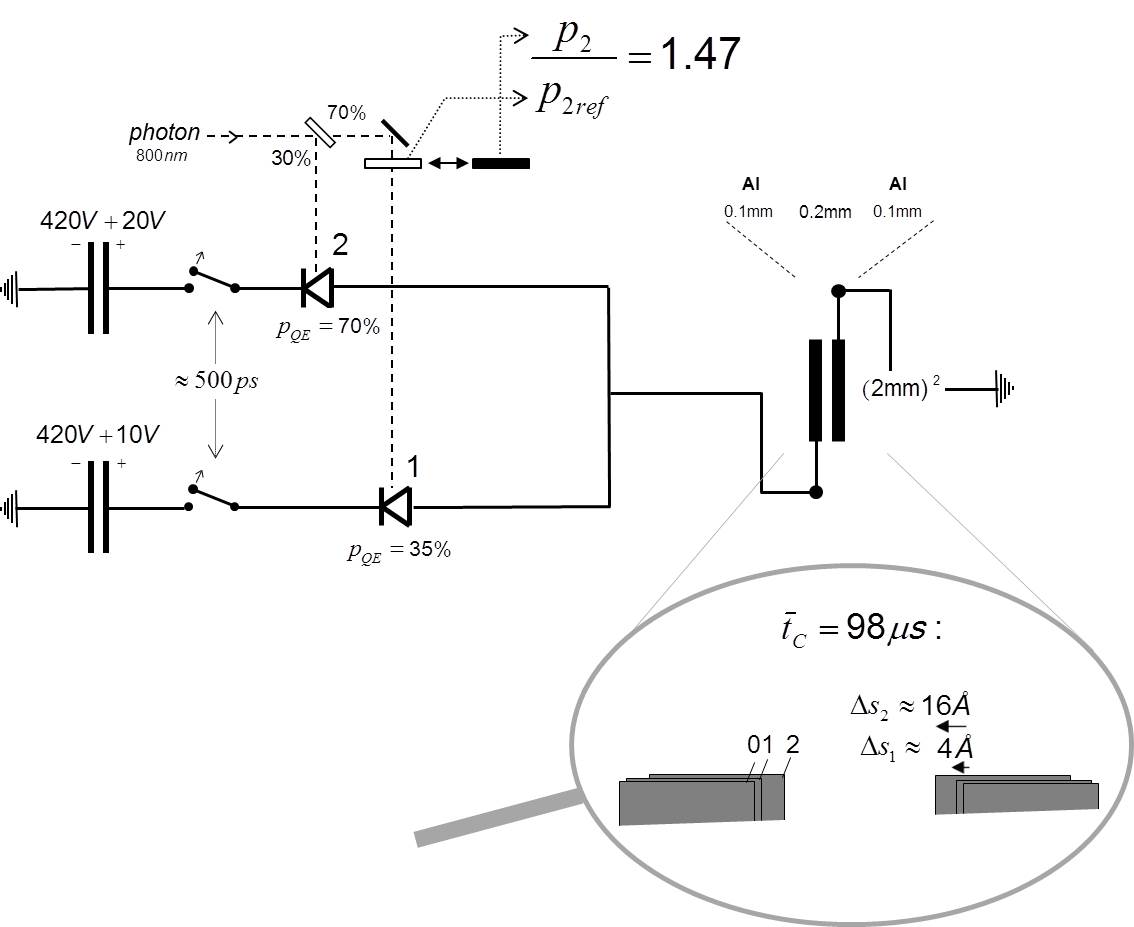}\vspace{0cm}
\caption{\small
Chosen parameters and calculated result for a setup using the capacitor with movable plates in Figure \ref{fig7} to allow a solid evolve into a three-state superposition. The detection probability in Photodiode 2 increases by a factor of $1.47$ when the aperture before Photodiode 1 is removed. The inset shows the displacements of the plates in States 1 and 2 at the reduction point in time $\bar{t}_{_{C}}$.
}
\label{fig8}
\end{figure}

\bigskip   
\bigskip
\bigskip
\noindent
\textbf{Choice of parameters} \\  
\noindent
From Equation (\ref{eq:38}), it follows that the displacement between States 0 and 2 at the reduction point in time $\Delta s_{2}(\bar{t}_{_{C}} )$ decreases with the mass density $\rho_{_{m}}$, the area $A$, the thicknesses $d_{_{m}}$ of the plates, and also with the distance $d$ between the plates. This leads us to choose aluminium plates with a small mass density $\rho_{_{m}}$, and to a small capacitor size: $A$$=$$(2mm)^{2}$, $d_{_{m}}$$=$$0.1mm$ and $d$$=$$0.2mm$, as displayed in Figure \ref{fig8}. The excess bias voltages of the photodiodes are chosen as $V_{_{E1}}$$=$$\,10V$ and $V_{_{E2}}$$=$$\,20V$, which leads, according to Equation (\ref{eq:36}), to a displacement ratio of $\Delta s_{2}/\Delta s_{1}$$=$$4$ at all times. A resistor behind Photodiode 1, as in the setup in Figure \ref{fig3}, is not needed.

\bigskip   
\bigskip
\bigskip
\noindent
\textbf{Feasibility} \\  
\noindent
For our choice of parameters, we obtain with Equations (\ref{eq:37}) and (\ref{eq:38}) a reduction point in time of $\bar{t}_{_{C}}$$\approx$$\,96\mu s$ and a displacement between States 0 an 2 at this point in time of $\Delta s_{2}(\bar{t}_{_{C}} )$$\approx$$15${\footnotesize \AA}. The increase of State 2's reduction probability follows with Equation (\ref{eq:21}) and $I_{_{2}}$$=$$0.21$ to be $p_{_{2}}/I_{_{2}}$$=$$1.65$. The exact procedure for calculating $\bar{t}_{_{C}}$ and $p_{_{2}}$ according to Section \ref{sec:3.4}, taking the short-distance contribution to the Di\'{o}si-Penrose energies of the plates into account, gives $\bar{t}_{_{C}}$$=$$\,98\mu s$, $p_{_{2}}/I_{_{2}}$$=$$1.47$, $\Delta s_{2}(\bar{t}_{_{C}} )$$=$$15.8${\footnotesize \AA} and $\Delta s_{1}(\bar{t}_{_{C}} )$$=$$3.95${\footnotesize \AA}, as shown in Figure \ref{fig8}. The displacement between States 0 and 1 at the reduction point in time of $\Delta s_{1}(\bar{t}_{_{C}} )$$=$$3.95${\footnotesize \AA} is still sufficiently large to fulfil the decorrelation criterion {\small$\Delta s_{1}(\bar{t}_{_{C}} )$$>$$6 \sigma_{n}$} ($\sigma_{n}$$=$\,$0.1${\footnotesize \AA} for aluminium). 

\bigskip
\noindent
The time constant for charging the capacitor with movable plates is $\tau$$=$$\,89ps$\footnote{\small   
$\tau$$=$$R_{_{d}}C$ with $R_{_{d}}$$=$$500\Omega$ and  $C$$=$$\epsilon_{_{0}}A/d$.
}, significantly smaller than the reduction point in time of $\bar{t}_{_{C}}$$\approx$$\,100\mu s$, which justifies the approach $V_{_{i}}(t)$$=$$V_{_{Ei}}$ for the voltage profiles. The short charging time allows us to disconnect the photodiodes from their biasing plate capacitors after $500ps$, as shown in Figure \ref{fig8}, which suppresses the dark count probability $p_{_{DC}}$ to a minimum. 

\bigskip
\noindent
The chosen parameters and calculated results for the experiment with the movable plates capacitor are displayed in Figure \ref{fig8}.

\newpage
%
%
%
\subsection{Impact of the other components}                 
%
%
\label{sec:5.3}
In this section, we investigate the influence of the other components of the setup on the reduction point in time, such as the photodiodes, the biasing plate capacitors, etc., and give recommendations on how to choose them. For the biasing plate capacitors, it is recommended not to use electrolytic capacitors, but instead simple plate capacitors with dielectrics of corundum; and for the resistor $R$ behind Photodiode 1 not commercially available metal-film resistors, but instead a rod of doped silicon.

\bigskip   
\bigskip
\bigskip
\noindent
\textbf{Characteristic lifetime $T_{_{G}}$ of a component} \\  
\noindent
The other components of our setup that are used to let the piezo capacitor evolve into a three-state superposition, such as the photodiodes, the biasing plate capacitors, the resistor behind Photodiode 1, etc., belong in the thought experiment in Figure \ref{fig1} to one of the two detectors, whose Di\'{o}si-Penrose energies are described by Equation (\ref{eq:3}), where the Di\'{o}si-Penrose energies $E_{_{Gi}}$ of a component $i$ depend on the mass distributions in the detection $\rho_{_{det}}(\mathbf{x})$ and no-detection states $\rho_{_{no-det}}(\mathbf{x})$ of the component (cf. Equation \ref{eq:3}). To simply our discussion, we ignore the time-dependencies of these Di\'{o}si-Penrose energies, and calculate the Di\'{o}si-Penrose energy $E_{_{Gi}}$ for every component after the settling time of detection. With the Di\'{o}si-Penrose criterion $T_{_{G}}$$=$$\,\hbar/E_{_{G}}$, we can characterise every component by a characteristic lifetime $T_{_{Gi}}$ that corresponds to the lifetime of the two-state superposition of the detection and no-detection states after the settling of this component ($T_{_{Gi}}$$=$$\,\hbar/E_{_{Gi}}$). 

\bigskip
\noindent
The contribution of Detectors 1 and 2 to the total competition action at the reduction point in time in Equation (\ref{eq:14}) $S^{^{Di}}_{G}(\bar{t}_{_{C}})$ can be calculated with the characteristic lifetimes of its components $T_{_{Gi}}$ by

\begin{equation}
\label{eq:39}
S^{^{D}}_{G}(\bar{t}_{_{C}}) \approx
\hbar \sum_{i} \frac{\bar{t}_{_{C}}}{T_{_{Gi}}}
\textrm{\textsf{~~.~~~~~~~~~~~~~~~~~}}
\end{equation}

~
\newline
\noindent When the characteristic lifetimes of the components $T_{_{Gi}}$ are significantly larger than the reduction point in time $\bar{t}_{_{C}}$ ($T_{_{Gi}}$$>>$$\bar{t}_{_{C}}$), their impacts on $\bar{t}_{_{C}}$ are negligible.

\bigskip   
\bigskip
\bigskip
\noindent
\textbf{Calculation of the components' lifetimes in \cite{Solid}} \\  
\noindent
The calculation of the components' characteristic lifetimes $T_{_{Gi}}$ requires a detailed analysis of all physical processes that can change the mass distribution $\rho_{_{det}}(\mathbf{x})$ in the detection state of the component. This analysis is carried out in \cite{Solid}, in which the lifetime of a single-photon detector is calculated, which consists of the same components as our setup. The calculations in \cite{Solid} are therefore, apart from the chosen parameters, identical to ours. Hence, we restrict ourselves here to the results of these calculations, and refer the reader for further details to \cite{Solid}. 

\bigskip
\noindent
The calculations in \cite{Solid} show that the displacements between the nuclei in the detection and no-detection states of the component are much smaller than the nuclei's spatial variations of about $\sigma_{n}$$\approx$\,$0.1${\footnotesize \AA}. In this limiting case, the Di\'{o}si-Penrose energy of the component is dominated (according to the discussion in Section \ref{sec:3.3}) by the short-distance contribution, which can be calculated with Equation (\ref{eq:13}) for $\Delta s$$<<$$\sigma_{n}$.

\bigskip   
\bigskip
\bigskip
\noindent
\textbf{Plate capacitors} \\  
\noindent
The voltages at the biasing plate capacitors are slightly decreased when they have finished charging the piezo capacitor. Due to the electric forces between the capacitor's plates, its dielectric is compressed, and we have in the no-detection state with a voltage of $V_{_{B}}$$+$$V_{_{E}}$ a slightly larger compression than in the detection state with a voltage of $V_{_{B}}$$+$$V_{_{E}}$$-$$\Delta V$ ($\Delta V$$\approx$$V_{_{E}}\cdot C_{_{p}}/(C$$+$$C_{_{p}})$; cf. Section \ref{sec:4.3}). The formula for the Di\'{o}si-Penrose energy of a plate capacitor in a two-state superposition with different voltages is derived in \cite{Solid}. To obtain a large characteristic lifetime $T_{_{G}}$ of the plate capacitor, it is recommended to use dielectrics of corundum ($Al_{_{2}}O_{_{3}}$), which has a modulus of elasticity of about $E_{_{e}}$$\approx$$\, 350$$-$$\, 406GPa$
 \cite{Gen-12}.

\bigskip
\noindent
For the voltage profiles in Equation (\ref{eq:29}), it is assumed that the capacitances of the plate capacitors $C$ are much larger than that of the piezo capacitor $C_{_{p}}$ ($C$$>>$$C_{_{p}}$).  The piezo capacitor in Figure \ref{fig6} has a capacitance of $C_{_{p}}$$=$$1300pF$. For a plate capacitor with an area of $(9cm)^{2}$ and a thickness of $1mm$, we obtain with $\epsilon_{_{r}}$$\approx$$9$ for corundum \cite{Gen-12} a capacitance of $C$$\approx$$650pF$. To satisfy $C$$>>$$C_{_{p}}$, we have to connect approximately 18 of such capacitors in parallel. The voltage for these 18 plate capacitors in parallel for biasing Photodiode 2 will decrease from $420V$$+$$20V$ to approximately $420V$$+$$18V$ after the piezo capacitor is charged. With the formula for the Di\'{o}si-Penrose energy of a plate capacitor in a two-state superposition in \cite{Solid}, we obtain a characteristic lifetime of the 18 plate capacitors in parallel of $T_{_{G}}$$\approx$$\,100ms$, where plates of copper with thicknesses of $0.03mm$ are assumed.

\bigskip
\noindent
When the capacitances of the biasing plate capacitors are not significantly larger than that of the piezo capacitor ($C$$>>$$C_{_{p}}$), one has to make the substitutions $V_{_{Ei}}$$\rightarrow$$\alpha V_{_{Ei}}$ and $C_{_{p}}$$\rightarrow$$\alpha C_{_{p}}$ with $\alpha$$=$$C/(C$$+$$C_{_{p}})$ in Equation (\ref{eq:29}). This leads to the substitutions $V_{_{E2}}$$\rightarrow$$\alpha V_{_{E2}}$ and $\epsilon_{_{r}}$$\rightarrow$$\alpha \epsilon_{_{r}}$ in Equations (\ref{eq:33})-(\ref{eq:35}) for estimating the reduction point in time and the displacement at this point in time.

\bigskip   
\bigskip
\bigskip
\noindent
\textbf{Resistor, photodiodes, wires and switches} \\  
\noindent
The photodiodes, the resistor, the wires (connecting the components) and the two switches in Figure \ref{fig6} will have in the detection state a slightly larger thermal expansion due to the heat energy that is generated by the avalanche current in photon detection. The formulae for the Di\'{o}si-Penrose energies of these components in two-state superpositions with a slightly larger thermal expansion in the detection than in the no-detection state are derived in \cite{Solid}. The heat energy that is generated by the avalanche current is given by $R\int^{t_{q}}_{0}$$dtI(t)^{2}$, where $I(t)$ is the avalanche current's profile, $t_{_{q}}$ the point in time at which the avalanche breaks down, and $R$ the resistance of the component. In the calculations of the following results, we used $\int^{t_{q}}_{0}$$dtI(t)^{2}$$\approx$$\,500(mA)^{2}\mu s$ following from the avalanche current profile $I(t)$ in Photodiode 2 in the experiment in Figure \ref{fig6}, which breaks down at $t_{_{q}}$$\approx$$\,4\mu s$ (cf. Section \ref{sec:5.1.3}).

\bigskip
\noindent
\textbf{Resistor:} In \cite{Solid}, it is shown that the Di\'{o}si-Penrose energy of a resistor depends on the resistivity $\rho_{_{\Omega}}$ of the used material as $E_{_{G}}$$\propto$$\rho^{^{-1}}_{\Omega}$. To obtain large characteristic lifetimes $T_{_{G}}$, it is recommended not to use commercially available metal-film resistors with resistivities of about $\rho_{_{\Omega}}$$\approx$$10^{-6}\Omega cm$, but instead a rod of doped silicon, with a much higher resistivity of e.g. $\rho_{_{\Omega}}$$\approx$$\,5\Omega cm$ corresponding to n-type doped silicon with a doping concentration of roughly $10^{15}/cm^{3}$ \cite{Gen-12}. The $940 \Omega$-resistor in Figure \ref{fig6} can then by realised by a rod of diameter of $2mm$ and a length of approximately $6cm$. With the formulae derived in \cite{Solid}, we obtain a characteristic lifetime of $T_{_{G}}$$\approx$$\,1s$ for this component.

\bigskip
\noindent
\textbf{Photodiodes:} The adaption of the calculation of the characteristic lifetime of the photodiode of the single photon detector in \cite{Solid} yields, for our avalanche current profile of $\int^{t_{q}}_{0}$$dtI(t)^{2}$$\approx$$\,500(mA)^{2}\mu s$, a characteristic lifetime of about $T_{_{G}}$$\approx$$\,1s$.

\bigskip
\noindent
\textbf{Wires:} For the wires connecting the components of our setup in Figure \ref{fig6}, we regard representative a wire of copper with a length of $200cm$, a diameter of $1mm$, and an effective length of $l_{_{e}}$$=$$4cm$, where the effective length $l_{_{e}}$ is defined in \cite{Solid}. With the formulae in \cite{Solid}, we obtain a characteristic lifetime of about $T_{_{G}}$$\approx$$\,10^{7}s$. 

\bigskip
\noindent
\textbf{Switches:} The characteristic lifetimes of the switches in Figure \ref{fig6} for disconnecting the photodiodes from the biasing plate capacitors after photon detection, which can be realised by field effect transistors of depletion mode, are expected to be in between those of the $940 \Omega$-resistor ($T_{_{G}}$$\approx$$\,1s$) and the wires ($T_{_{G}}$$\approx$$\,10^{7}s$).

\newpage
%
%
%
\section{Pursuing experiments}                 
%
%
\label{sec:6}
In this section, we take a look on pursuing experiments that were proposed in \cite{P2,NS} to check further aspects of the Dynamical Spacetime approach. For these experiments, we propose concrete setups and show their feasibility. In Section \ref{sec:6.1}, we discuss an experiment for measuring the lifetime of a two-state superposition. In Section \ref{sec:6.2}, we discuss the signalling experiment proposed in \cite{P2,NS} for checking the quasi-abrupt reconfigurations of the wavefunction's evolution in the Dynamical Spacetime approach, which can cover far-separated regions.
\bigskip

%
\begin{figure}[h]
\centering
\includegraphics[width=10.5cm]{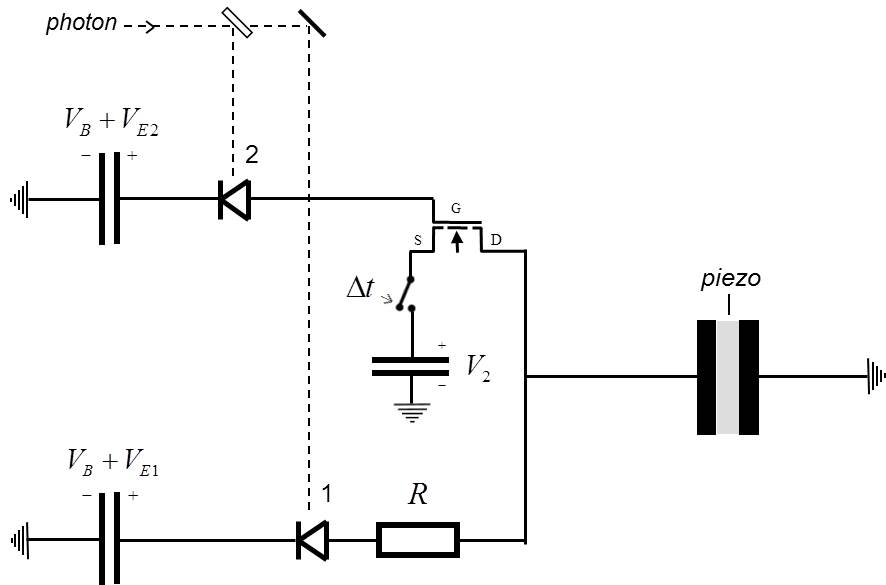}\vspace{0cm}
\caption{\small
Modification of the setup in Figure \ref{fig3} to measure the lifetime of the two-state superposition of States 0 and 1.
}
\label{fig9}
\end{figure}

%
%
%
\subsection{Lifetime of a two-state superposition}                 
%
%
\label{sec:6.1}
In \cite{P2,NS}, an experiment is proposed for measuring the lifetime of a two-state superposition with the help of the deviations from Born's rule predicted by the Dynamical Spacetime approach. If one allows the solid in Figure \ref{fig1} first to evolve into a two-state superposition by delaying the displacement $\Delta s_{1}$ between States 0 and 1 by a time delay of $\Delta t$, one observes an increased reduction probability of State 2 only, when the two-state superposition of States 0 and 2 has not reduced before $\Delta t$. By measuring the reduction probability of State 2 over the time delay $\Delta t$, one can determine the lifetime of the two-state superposition. This procedure allows one not only to check whether the reduction point in time $\bar{t}_{_{C}}$ follows the Di\'{o}si-Penrose criterion $\bar{t}_{_{C}}$$\approx$$\,\hbar/E_{_{G}}$, but also to check the most important prediction of the Dynamical Spacetime approach: that collapse is not possible at any point in time (as in dynamical reduction models \cite{GRW_Ue-2}), but only at specified points in time: the reduction points in time $\bar{t}_{_{C}}$. The reduction probability of State 2 over $\Delta t$ must therefore decrease abruptly at the reduction point in time $\bar{t}_{_{C}}$ of the two-state superposition.

\bigskip
\noindent
Figure \ref{fig9} shows how the setup in Figure \ref{fig3} has to be modified for this experiment. Different to that described above, we measure the lifetime of the two-state superposition of States 0 and 1 and delay the displacement $\Delta s_{2}$ between States 0 and 2 by $\Delta t$. This is realised by allowing the avalanche current of Photodiode 2 open the gate of a field effect transistor of enhancement mode, which connects the piezo capacitor with the plate capacitor with the voltage $V_{_{2}}$ in Figure \ref{fig9}, which is used for charging. The time delay is realised with the switch between the plate and the piezo capacitor, which is closed after the time delay $\Delta t$, as shown in Figure \ref{fig9}. 

\bigskip
\noindent
The lifetime, respectively the reduction point in time $\bar{t}_{_{C}}$, of the two-state superposition of States 0 and 1 can be varied by the choice of the resistance $R$ behind Photodiode 1, where one has to take care that the displacement between States 0 and 1 is (at the reduction point in time) still sufficiently large to fulfil the decorrelation criterion {\small$\Delta s_{1}(\bar{t}_{_{C}} )$$>$$6 \sigma_{n}$}. The voltage $V_{_{2}}$ of the plate capacitor for charging the piezo capacitor has to be chosen in such a way that the displacement $\Delta s_{2}$$=$$d_{_{33}}V_{_{2}}/2$, which is achieved when the switch is closed (cf. Equation \ref{eq:30}), is at least four times larger than that between States 0 and 1 ($\Delta s_{2}$$>$$4\Delta s_{1}(\bar{t}_{_{C}})$) to observe a significant increase of State 2's reduction probability for $\Delta t$$<$$\bar{t}_{_{C}}$.

\bigskip
%
%
%
\subsection{Signalling experiment}                 
%
%
\label{sec:6.2}
In \cite{P2,NS}, it is shown that the deviations from Born's rule in the Dynamical Spacetime approach lead to superluminal signalling. This does not lead to a conflict with relativity, since causality evolves in the Dynamical Spacetime approach not along free selectable Lorentz frames in spacetime, but is parametrised by its expansion and evolves quasi-orthogonal to it \cite{P2,NS}.
\bigskip

%
\begin{figure}[h]
\centering
\includegraphics[width=16cm]{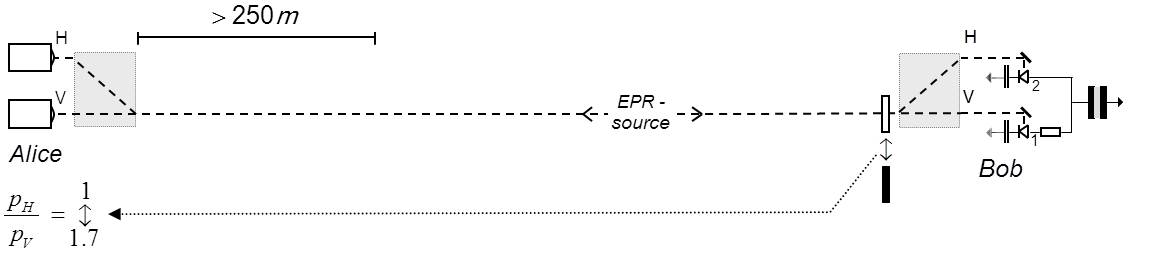}\vspace{0cm}
\caption{\small
Signalling experiment by inserting the setup of Figure \ref{fig6} into an EPR experiment. Bob can manipulate the ratio between the polarisation probabilities measured by Alice from usually $p_{_{H}}/ p_{_{V}}$$=$$1$ to $p_{_{H}}/ p_{_{V}}$$\approx$$1.7$ by removing the aperture.
}
\label{fig10}
\end{figure}

\noindent
A signalling experiment can be constructed by inserting the setup of Figure \ref{fig6} into an EPR experiment, as shown in Figure \ref{fig10}. The setup puts Bob in a position to change the ratio between the polarisation probabilities measured by Alice from usually $p_{_{H}}/ p_{_{V}}$$=$$1$ to $p_{_{H}}/ p_{_{V}}$$\approx$$1.7$, when he removes the aperture before his apparatus. This enables him to signal information to Alice. When the aperture is removed, the piezo capacitor on Bob's side evolves into a three-state superposition, which increases for a Bell state of $|\psi$$>$$=$$|H$$>$$|H$$>$$+|V$$>$$|V$$>$ the reduction probability of the $|H$$>$$|H$$>$-part. Since the quasi-abrupt reconfiguration of the wavefunction's evolution also covers Alice's side, the polarisation of Alice's photon changes instantaneously, according to the reduction in the piezo capacitor. This happens before the photon arrives at Alice's detectors, when her arm is chosen to be at least 

\begin{equation}
\label{eq:40}
\delta > c \bar{t}_{_{C}}
\end{equation}

~
\newline
\noindent longer than Bob's arm. For the reduction point in time of our setup of $\bar{t}_{_{C}}$$=$$\,0.84\mu s$, we obtain $\delta$$>$$250m$, as shown in Figure \ref{fig10}. Photons with a wavelength of approximately $\lambda$$\approx$$800nm$ for our setup in Figure \ref{fig6} can be generated by the usual parametric down-conversion of photons with wavelengths of $\lambda$$=$$404nm$. The ratio $p_{_{H}}/ p_{_{V}}$ between the polarisation probabilities (when the aperture is removed) can be calculated with Equation (\ref{eq:21}) as:\footnote{\small   
From Equation (\ref{eq:21})  for the case "$\Delta s_{2}$$>>$$\Delta s_{1}$" in Section \ref{sec:3.4}, it follows that the reduction probabilities of States 0, 1 and 2 are given by:
\newline
\newline
\hspace*{10mm}  $p_{_{0}}$$=$$\frac{I_{_{0}}}{1+I_{_{2}}}$, $p_{_{1}}$$=$$\frac{I_{_{1}}}{1+I_{_{2}}}$, $p_{_{2}}$$=$$\frac{2I_{_{2}}}{1+I_{_{2}}}$.
\newline
\newline
With this result, the probabilities $p_{_{H2}}$, $p_{_{H0}}$, $p_{_{V1}}$  and $p_{_{V0}}$ change as: 
\newline
\newline
\hspace*{10mm} $p'_{_{H2}}$$=$$\frac{2p_{_{H2}}}{1+p_{_{H2}}}$, $p'_{_{H0}}$$=$$\frac{p_{_{H0}}}{1+p_{_{H2}}}$, $p'_{_{V1}}$$=$$\frac{p_{_{V1}}}{1+p_{_{H2}}}$, $p'_{_{V0}}$$=$$\frac{p_{_{V0}}}{1+p_{_{H2}}}$,
\newline
\newline
where $p_{_{H2}}$ refers to the case that Alice measures an H-polarisation and Bob a photon detection by photodiode 2, and $p_{_{H0}}$ to the case that Bob measures no photon etc.
With $p_{_{H}}$$=$$p_{_{H2}}$$+$$p_{_{H0}}$, $p_{_{V}}$$=$$p_{_{V1}}$$+$$p_{_{V0}}$, $p'_{_{H}}$$=$$p'_{_{H2}}$$+$$p'_{_{H0}}$, $p'_{_{V}}$$=$$p'_{_{V1}}$$+$$p'_{_{V0}}$
 and $p_{_{H}}$$=$$p_{_{V}}$$=$$\frac{1}{2}$,  we obtain $p'_{_{H}}/p'_{_{V}}$$=$$1$$+$$2p_{_{H2}}$, which leads with $p_{_{H2}}$$=$$\frac{1}{2}p_{_{QE2}}$ to Equation (\ref{eq:41}).
}

\begin{equation}
\label{eq:41}
\frac{p_{_{H}}}{p_{_{V}}} \approx 1 + p_{_{QE2}}
\textrm{\textsf{~~.~~~~~~~~~~~~~~~~~}}
\end{equation}

~
\newline
\noindent This leads, for the quantum efficiency of Photodiode 2 in Figure \ref{fig6} of $p_{_{QE2}}$$=$$70\%$, to $p_{_{H}}/ p_{_{V}}$$\approx$$1.7$, as shown in Figure \ref{fig10}.

\newpage
%
%
%
\section{Discussion}                 
%
%
\label{sec:7}
Our study has shown that the deviations from Born's rule predicted by the Dynamical Spacetime approach can be verified by quite simple experiments, which mostly use commercially available single photon and piezo technology; and that the experiments can be performed at room temperature. Our study has shown that all components of the setup must be chosen carefully, and that their influences on the reduction point in time must be calculated. Furthermore, one needs good techniques to minimise the setup's interaction with the environment during superposition. 

\bigskip
\noindent
The fact that the observation of deviations from Born's rule requires a specially designed experiment explains why such deviations have not yet become conspicuous. The two discussed realisations of the experiment with the piezo capacitor and the capacitor with movable plates predict quite different reduction points in time (i.e. $\bar{t}_{_{C}}$$\approx$$\,1\mu s$ and $\bar{t}_{_{C}}$$\approx$$\,100\mu s$), but they have in common that the solid that evolves into the three-state superposition must be fairly small, with a volume on the order of one cubic millimetre.

\bigskip
\noindent
Our numerical calculations have shown that the approximation formulae derived here are sufficient for the dimensioning of the experiment. These formulae are Equations (\ref{eq:21}) and (\ref{eq:27}) for the reduction probability of State 2, and Equations (\ref{eq:33})-(\ref{eq:35}), (\ref{eq:37}) and (\ref{eq:38}) for the reduction point in time and the displacement at this point in time of the piezo and the movable plates capacitor. For the dimensioning of the experiment, the experimenter must not perform numerical calculations. The influences of the other components of the setup on the result (such as the photodiodes etc.) can be estimated with the formulae derived in \cite{Solid}. 

\bigskip
\noindent
Critical to the success of the experiment is the minimisation of environmental interaction during superposition to avoid a shortening of the reduction point in time. The switches for disconnecting the voltage supply and the voltmeter from the setup during superposition in Figure \ref{fig4} can be realised by field effect transistors in enhancement mode. Any possible remaining interactions of such electric circuits with the environment should be analysed in detail.  

\bigskip
\noindent
It is exciting whether we will obtain new results with the proposed experiments, shedding more light on the riddle of wavefunction collapse.

\bigskip   
\bigskip
\bigskip
\bigskip
\noindent
\textbf{\small Acknowledgements} \\  
\noindent
{\small
I would like to thank my friend Christoph Lamm for supporting me and for proofreading the manuscript.
}

%
%
%

\end{document}